\documentclass[prb,twocolumn,reprint,amsmath,amssymb,superscriptaddress,floatfix,longbibliography]{revtex4-1}
\usepackage[breaklinks=true,colorlinks,citecolor=blue,linkcolor=blue,urlcolor=blue]{hyperref}
\usepackage{float}
\usepackage{multirow}
\usepackage{array}
\usepackage{tabularx}

\usepackage{epsfig,mathrsfs,color,latexsym,subfigure,marginnote,graphicx,verbatim,relsize,mathrsfs,color,array,amsmath,amsfonts,amssymb,graphicx}
\def\be{\begin{equation}}
\def\ee{\end{equation}}

\def\bea{\begin{eqnarray}}
\def\eea{\end{eqnarray}}

\newcommand{\T}{\mathcal{T}}
\newcommand{\K}{\mathcal{K}}
\newcommand{\I}{\mathcal{I}}
\newcommand{\Cz}{\mathcal{C}_{3z}}
\newcommand{\Mx}{\mathcal{M}_1}

\newcommand{\kb}{\mathbf{k}}
\newcommand{\M}{\mathcal{M}}

\begin{document}
\title{Saddle-point van Hove singularity and dual topological state in Pt$_2$HgSe$_3$ }

\author{Barun Ghosh}
\affiliation{Department of Physics, Indian Institute of Technology Kanpur, Kanpur 208016, India}

\author{Sougata Mardanya}
\affiliation{Department of Physics, Indian Institute of Technology Kanpur, Kanpur 208016, India}

\author{Bahadur Singh$^*$}
\affiliation{SZU-NUS Collaborative Center and International Collaborative Laboratory of 2D Materials for Optoelectronic Science $\&$ Technology, Engineering Technology Research Center for 2D Materials Information Functional Devices and Systems of Guangdong Province, College of Optoelectronic Engineering, Shenzhen University, ShenZhen 518060, China}
\affiliation{Department of Physics, Northeastern University, Boston, Massachusetts 02115, USA}

\author{Xiaoting Zhou}
\affiliation{Department of Physics and Astronomy, California State University, Northridge, CA 91330, USA}
\affiliation{Department of Physics, National Cheng Kung University, Tainan 701, Taiwan} 

\author{Baokai Wang}
\affiliation{Department of Physics, Northeastern University, Boston, Massachusetts 02115, USA}

\author{Tay-Rong Chang}
\affiliation{Department of Physics, National Cheng Kung University, Tainan 701, Taiwan} 
\affiliation{Center for Quantum Frontiers of Research and Technology (QFort), Tainan 701, Taiwan}

\author{Chenliang Su}
\affiliation{SZU-NUS Collaborative Center and International Collaborative Laboratory of 2D Materials for Optoelectronic Science $\&$ Technology, Engineering Technology Research Center for 2D Materials Information Functional Devices and Systems of Guangdong Province, College of Optoelectronic Engineering, Shenzhen University, ShenZhen 518060, China}

\author{Hsin Lin$^*$}
\affiliation{Institute of Physics, Academia Sinica, Taipei 11529, Taiwan}

\author{Amit Agarwal$^*$}
\affiliation{Department of Physics, Indian Institute of Technology Kanpur, Kanpur 208016, India}

\author{Arun Bansil\footnote{Corresponding authors' emails: bahadursingh24@gmail.com, nilnish@gmail.com, amitag@iitk.ac.in, ar.bansil@northeastern.edu}}
\affiliation{Department of Physics, Northeastern University, Boston, Massachusetts 02115, USA}

\begin{abstract}
Saddle-point van Hove singularities in the topological surface states are interesting because they can provide a new pathway for accessing exotic correlated phenomena in topological materials. Here, based on first-principles calculations combined with a $\mathbf {k \cdot p}$ model Hamiltonian analysis, we show that the layered platinum mineral jacutingaite (Pt$_2$HgSe$_3$) harbors saddle-like topological surface states with associated van Hove singularities. Pt$_2$HgSe$_3$ is shown to host two distinct types of nodal lines without spin-orbit coupling (SOC) which are protected by combined inversion ($\I$) and time-reversal ($\T$) symmetries.  Switching on the SOC gaps out the nodal lines and drives the system into a topological state with nonzero weak topological invariant $Z_2=(0;001)$ and mirror Chern number $n_M=-2$. Surface states on the naturally cleaved (001) surface are found to be nontrivial with a unique saddle-like energy dispersion with type II van Hove singularities. We also discuss how modulating the crystal structure can drive Pt$_2$HgSe$_3$ into a Dirac semimetal state with a pair of Dirac points. Our results indicate that Pt$_2$HgSe$_3$ is an ideal candidate material for exploring the properties of topological insulators with saddle-like surface states.
\end{abstract}

\maketitle

\section{Introduction}
Finding new topological materials with unique properties is currently drawing intense interest as an open research frontier in condensed matter physics and related fields.\cite{Bansil16,Hasan10,Qi11} Initial ideas of time-reversal symmetry ($\T$) protected topological states have been generalized to incorporate crystal symmetries, leading to the identification of a variety of new topological states in insulators, semimetals, and metals.\cite{Fu07,Fu_TCI,Po2017,Chen_Fang_SI,PhysRevX.7.041069} Examples include mirror-symmetry protected topological crystalline insulators (TCIs), weak topological insulators (WTIs), Dirac/Weyl semimetals, nodal line semimetals, hourglass semimetals, triple-point semimetals, among others.\cite{Hsieh2012,Kane_DSM_3D,DSM_WSM_review_viswanath,Na3Bi_PRB,Cd3As2_PRB,singh12,Z2NL_Fu,Z2NL_Kane,Fang_2016,Wang17,hourglass_Ag2BiO3,triplepoint_1}   Theoretically predicted topological properties of a number of materials have been demonstrated experimentally via spectroscopic and transport measurements.\cite{Tanaka2012,Xu2012,Neupane2014,Liu864,TaAs_expt,Xu613,Bian2016,Hourglass_expt} It has been recognized that a topological state can also be protected simultaneously by different crystal symmetries as is the case in Bi$_2$(Se,Te)$_3$ where the protection involves both $\T$ and crystalline mirror symmetries. \cite{Bi2Te3_dualtopology,Bi1Te1_dual} Such dual-symmetry-protected topological states can open up new possibilities for tuning topological properties via controlled symmetry breaking. 

Topological surface states (TSSs) are the hallmark and source of numerous useful properties in topological quantum materials. Depending on the symmetries of their crystalline surfaces, the electronic dispersion ($E_{\bf k}$) of TSSs can deviate substantially from the well-known Dirac-like form.\cite{singh18_saddle} Specifically, when a surface lacks rotational symmetry $C_{n}$ for $n>2$, a saddle-like $E_{\bf k}$ dispersion with saddle points is, in principle, allowed via symmetry constraints. Such saddle points in $\mathbf{k}$-space can lead to Van Hove singularities (VHSs) where densities of states (DOSs) diverge logarithmically in two-dimensions (2D).  The interest in VHSs has been revived recently in the theory of correlated twisted bilayer graphene and, in fact, the new concept of higher order VHSs has been proposed. \cite{TBLG_correlated_insulator,TBLG_superconductivity,LiangFu_higher_order_VHS} More generally, when VHSs lie close to the Fermi level, the increased DOS amplifies electron correlation effects that can drive various quantum many-body instabilities involving the lattice, charge and spin degrees of freedom.\cite{PhysRevLett.78.1343,PhysRevLett.87.187004,PhysRevB.92.085423,PhysRevB.95.245136,DSS_surface_FM} When these VHSs lie at generic $\kb$ points, they favor an odd-parity pairing, which can lead to unconventional superconductivity. \cite{VHS_superconductivity,MARKIEWICZ_VHS_SC} Despite theoretical prediction of TSSs with VHSs, experimental evidence of such states is still lacking. The identification of new materials with saddle-like TSSs is thus of great importance.

Here, we investigate the topological electronic structure of layered platinum mineral jacutingaite Pt$_2$HgSe$_3$ and reveal a dual-symmetry-based protection of its topological state and the existence of saddle-point VHSs in its surface electronic spectrum. The monolayer Pt$_2$HgSe$_3$ has been predicted recently as a large band gap Kane-Mele quantum spin Hall (QSH) insulator.\citep{HgPt2Se3_monolayer} A nontrivial band gap of 0.53 eV has been found within the G$_0$W$_0$ approximation: its Fermiology under electron and hole doping suggests the existence of VHSs and unconventional superconductivity.\cite{HgPt2Se3_superconductivity} The QSH state in  Pt$_2$HgSe$_3$ monolayer has been experimentally demonstrated using scanning tunneling microscopy (STM).\cite{HgPt2Se3_expt} Also, it is found that few nanometers thick as well as bulk jacutingaite is stable under ambient conditions for months and even up to an year.\cite{HgPt2Se3_expt} However, the bulk topological state and the associated TSSs with VHSs remain unexplored.  

Our analysis reveals that Pt$_2$HgSe$_3$ supports two distinct types of nodal lines when spin-orbit coupling (SOC) effects are ignored. Including SOC in the computations gaps out the nodal lines and drives the system into a topological state characterized by nonzero weak topological invariants, $Z_2=(0;001)$, as well as the mirror Chern number $n_M=-2$. To highlight the nontrivial bulk band topology, we investigate the naturally cleaved (001)-surface electronic structure and show the existence of a unique symmetry-allowed saddle-like $E_{\bf k}$ dispersion of topological surface states with saddle-point VHSs. Informed by our first-principles computations, we present a viable ${\bf k.p}$ model Hamiltonian for the topological surface states. We also discuss the effect of hydrostatic pressure on bulk band topology and reveal the presence of a topological phase transition to a type-II Dirac semimetal state with pressure. Our results suggest that Pt$_2$HgSe$_3$ is an ideal material for experimental exploration of saddle-like surface states with VHSs. 

The remainder of the paper is organized as follows. In Sec.~\ref{Comp_details}, we discuss computational details along with the crystal structure of Pt$_2$HgSe$_3$. The bulk topological properties are discussed in Sec.~\ref{bulkES}. In section~\ref{TSSs}, we characterize the topological states and present surface electronic structure with and without SOC. The ${\bf k.p}$ model Hamiltonian for the topological surface states is described in Sec.~\ref{modelTSSs}. In Sec.~\ref{TPTs}, we present the evolution of topological electronic structure under hydrostatic pressure. Finally, we summarize our findings in Sec.~\ref{con}. 

\section{Computational Details and Crystal Structure} \label{Comp_details}

Electronic structure calculations were performed within the framework of the density functional theory (DFT) with the projector-augmented-wave (PAW) pseudopotentials and a plane-wave basis set using the Quantum Espresso package.\cite{PhysRev.140.A1133,PhysRevB.59.1758,Giannozzi_2009} We used an energy cut-off of 50 Ry for the plane wave basis set and a $9 \times 9 \times 8$ $\kb$ mesh for the bulk calculations.  The generalized gradient approximation (GGA) of Perdew, Burke, and Ernzerhof (PBE) was used to include exchange-correlation effects.\cite{PhysRevLett.77.3865} A tolerance of 10$^{-8}$ Ry was used for electronic energy minimization.
{Experimental lattice parameters ($a=b=7.348$ \AA~ and $c=5.295$ \AA) were used, but the atomic positions within the unit cell were optimized until the residual force on each atom was less than $10^{-3}$ Ry/au; see Appendix \ref{appA} for structural details. Results presented in this study are based on the GGA-PBE. Effects of van der Waal's corrections,  which we ascertained using the DFT-D3 method\cite{GrimmieDFTD3BJ}, were found to be negligible.}
We constructed our tight-binding model Hamiltonian by deploying atom-centered Wannier functions and computed topological properties using the WannierTools package.\cite{PhysRevB.56.12847,WU2017} The surface electronic spectrum was also checked by calculations using a supercell of ten-layer thick Pt$_2$HgSe$_3$ slabs separated by vacuum regions of 16 \AA~ using the VASP suite of codes.\cite{PhysRevB.54.11169}$^,
\footnote{{Both the Quantum Espresso and VASP codes were used for calculating structural and electronic properties. Two sets of results were found to be in excellent agreement.}}$

\begin{figure}
\includegraphics[width=\linewidth]{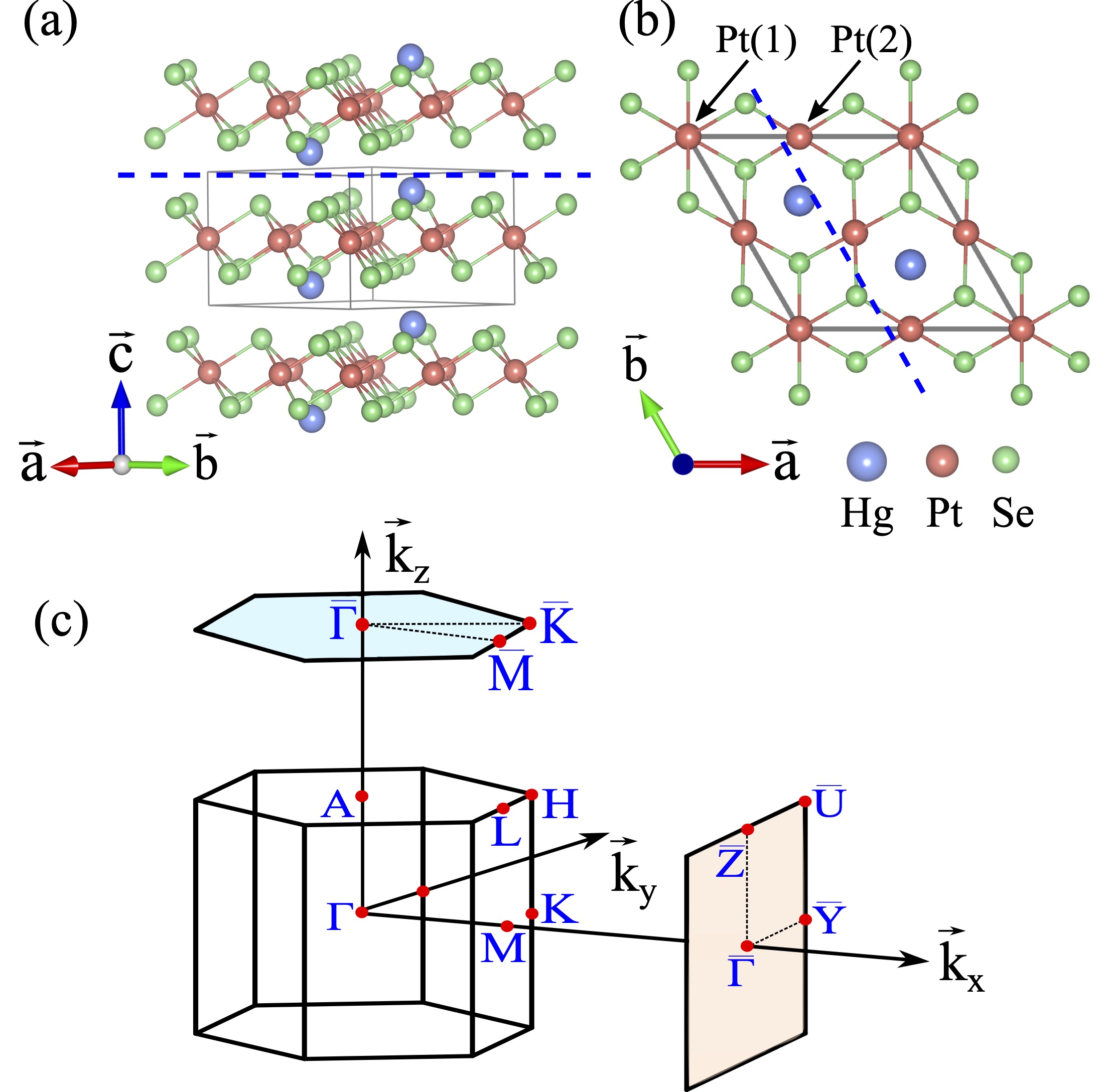}
\caption{(a) Side and (b) top view of the layered crystal structure of Pt$_2$HgSe$_3$. Pt(1) and Pt(2) denote two symmetry inequivalent Pt atoms in the unit cell. {The dashed-blue line in (a) indicates the natural surface termination along the (001)-surface direction.} (c) Bulk and projected (100) and (001) surface Brillouin zones. Various high-symmetry points are marked.}
\label{fig1}
\end{figure}

Jacutingaite Pt$_2$HgSe$_3$ forms a bipartite lattice in the {symmorphic} space group $P\bar{3}m1$ (No. 164).\cite{HgPt2Se3_structure_1} {The experimental crystal structure is layered with AA stacking} and it can be viewed as a $2\times2\times1$ supercell of 1T-PtSe$_2$ with additional Hg atoms that are placed in the anti-cubo-octahedral voids of Se atoms [Figs. \ref{fig1}(a)-(b)]. There are two symmetry-inequivalent Pt atoms in the primitive unit cell that form two distinct hexagonal sublattices. The Pt(1) atom connects to six nearest Se atoms and forms Pt(1)Se$_6$ local octahedral coordination while the Pt(2) atom constitutes the Pt(2)Se$_4$ square structure.\citep{HgPt2Se3_structure_1,HgPt2Se3_structure_2} The Pt(1)-Se bond length is 2.55 \AA, which is slightly larger than the Pt(2)-Se bond length of 2.47 \AA. {This crystal structure possesses three-fold rotational symmetry around the $z$-axis ($C_{3z}$), inversion symmetry $\I$, and the mirror symmetries $ \M_{100}$, $\M_{010}$ and $\M_{110}$}. Additionally, it respects the $\T$-symmetry.  

\section{Bulk electronic structure and topological invariants} \label{bulkES}

\begin{figure}[t!]
\includegraphics[width=\linewidth]{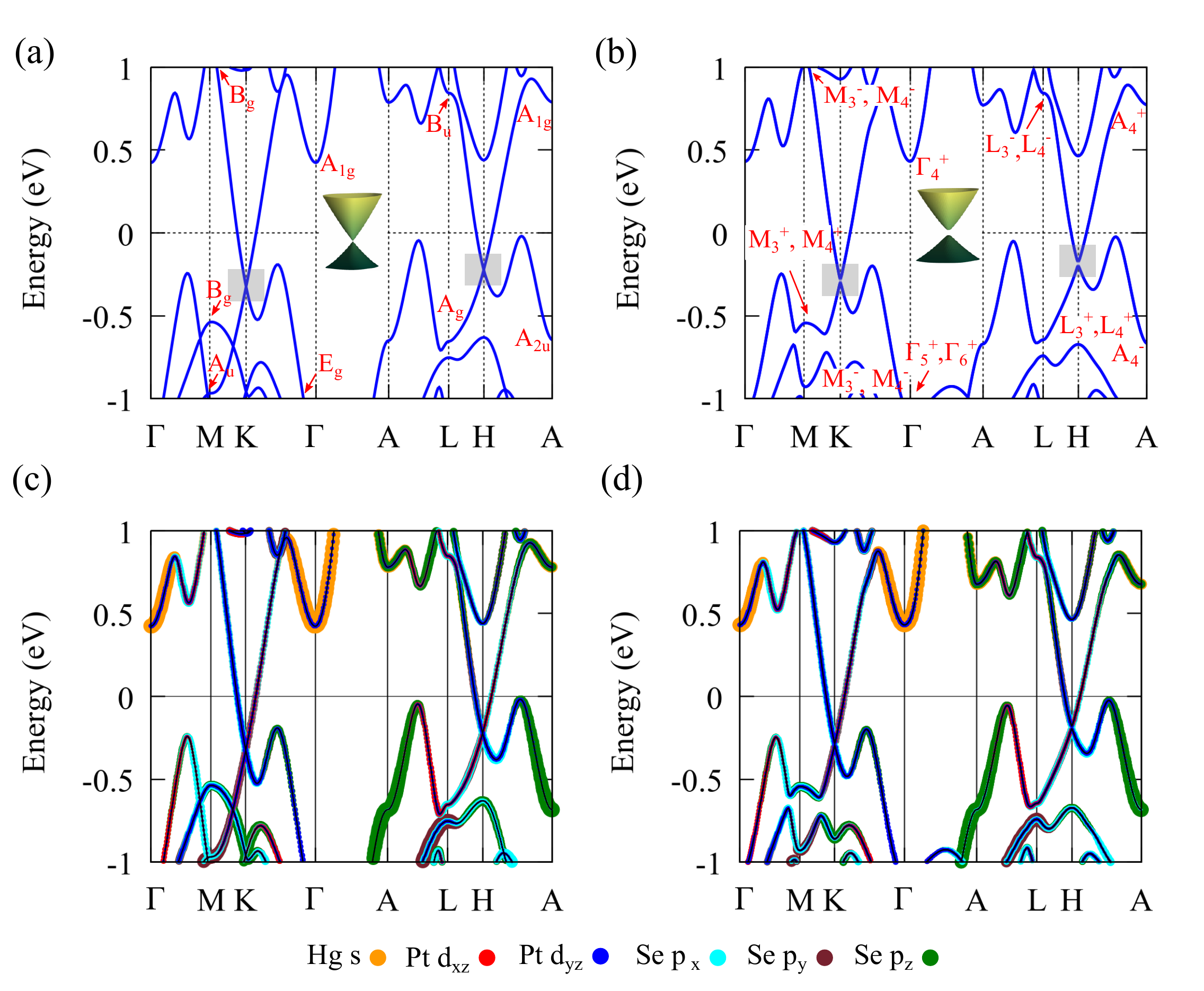}
\caption{Bulk band structure of Pt$_2$HgSe$_3$ (a) without and (b) with spin-orbit coupling (SOC). Irreducible representations at the time-reversal invariant momentum points are marked for bands near the Fermi energy. The insets in (a) and (b) show closeups of the in-plane energy dispersion near the $K$ and $H$ momentum points of the region highlighted by gray boxes in the main figure. A clear gap is seen to emerge at the $K$ and $H$ points with SOC in (b). (c) and (d) show the orbital-resolved band structure of Pt$_2$HgSe$_3$ without and with SOC, respectively.}
\label{fig2}
\end{figure}

The bulk electronic spectrum of Pt$_2$HgSe$_3$ (without SOC) is shown in Fig.~\ref{fig2}(a). It is semimetallic in character where the $A_u (A_g)$ symmetry band is seen to cross with the $B_g (B_u)$ band at the $K (H)$ point in the bulk BZ. These band crossings are linearly dispersed over a substantial energy range along ${\Gamma MK\Gamma}$  and ${ALHA}$. Similar Dirac-cone-like band features are also present in the band structure of graphite and their origin is attributed to the honeycomb lattice arrangement of the constituent atoms.\citep{Graphite_BS} The orbital-resolved band structure in Fig.~\ref{fig2}(c) shows that these crossing bands are mainly composed of Hg $s$, Se $p$, and Pt $d_{xz}$ and $d_{yz}$ orbitals. The band structure including the SOC is illustrated in Figs.~\ref{fig2}(b) and \ref{fig2}(d). The Dirac-cone-like band crossings without the SOC at the $K$ and $H$ points are now gapped and a continuous bandgap appears between the valence and conduction bands.

\begin{figure*}
\includegraphics[width=\linewidth]{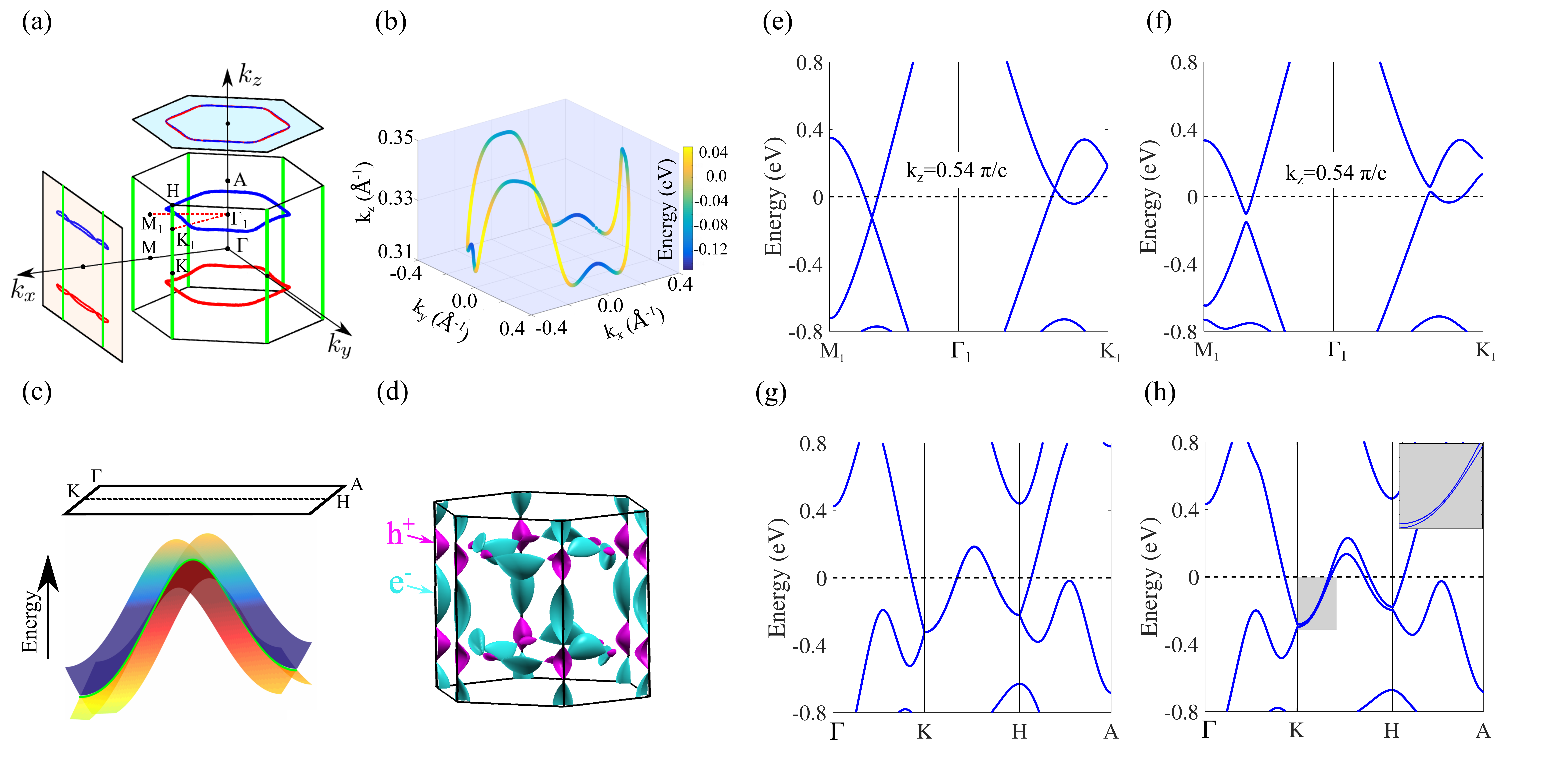}
\caption{(a) Nodal lines in the bulk BZ and their projections on the (100) and (001) surface planes in Pt$_2$HgSe$_3$. Two distinct types of nodal lines are shown. Type-I nodal lines (NL$_{\rm{C}}$) are located inside the bulk BZ and marked in red and blue colors. Type-II nodal lines (NL$_{\rm{KH}}$) are located along the hinges and shown in solid-green color. (b) Energy-momentum spread of the NL$_{\rm{C}}$ in the BZ.  (c) Schematics of NL$_{\rm{KH}}$ structure in the BZ. (d) Fermi surface with electron and hole pockets (without SOC). 
Band structure in the $k_z =0.54 (\frac{\pi}{c})$ plane (e) without and (f) with SOC. The nodal band crossings are shown resolved along $\Gamma_1-M_1$ and $\Gamma_1-K_1$ in (e). (g) and (h) show energy dispersion of the NL$_{\rm{KH}}$ nodal line without and with SOC, respectively. }
\label{fig3}
\end{figure*}

In order to characterize the nodal lines and their symmetry protection, we systematically examine the band crossings in Fig. \ref{fig3}. A careful inspection of band crossings in full bulk BZ reveals that Pt$_2$HgSe$_3$ hosts two distinct types of nodal lines. The type I nodal lines (identified by NL$_{C}$) are generated by accidental band crossings and form an inversion-symmetric pair of closed loops at generic $\kb$ points around the $\Gamma-A$ line inside the BZ. Importantly, these nodal lines are not hooked to a fixed momentum plane but trace an arbitrary path encircling the $\Gamma A$ line [see red and blue curves in Fig.~\ref{fig3}(a)]. They show considerable energy spread in the momentum space as illustrated in Fig.~\ref{fig3}(b) where the energies of the gap closing points are plotted in color in the $k_x-k_y-k_z$ momentum space. We further demonstrate these nodal crossings by plotting the band structure along the in-plane directions for a fixed $k_z =0.54 (\frac{\pi}{c})$ plane in Fig.~\ref{fig3}(e). 

The type II nodal lines (NL$_{\rm KH}$) stretch along the $K-H$ high symmetry directions at the hinges of the hexagonal BZ [green lines in Fig.~\ref{fig3}(a)]. These nodal lines are essential and enforced by little group symmetries of the $KH$ line. Notably, $KH$ line is invariant under three-fold rotational symmetry $\Cz$ and anti-unitary operator $\I \T$. For a spinless system, the eigenvalues of $\Cz$ are 1, $e^{+i\frac{2\pi}{3}}$, and $e^{-i\frac{2\pi}{3}}$. The conjugate symmetry operator $\I\T$, however, enforces a double degeneracy between states with $e^{i\frac{2\pi}{3}}$ and $e^{- i\frac{2\pi}{3}}$ eigenvalues. We have verified these symmetry states through an analysis of our first-principles wavefunctions. We find that the symmetry-adapted basis $\Psi ={(\psi_{+}, \psi_{-})}^T$ of the degenerate bands can be expressed as $ \psi_{\pm} = w_1 |p_x \pm ip_y\rangle + w_2|d_{xz}\pm id_{yz}\rangle +w_3|d_{x^2-y^2}\mp 2i d_{xy}\rangle $ where  $w_{i=1,2,3}$ are normalized coefficients. We further explore the nodal line energy dispersions in Figs. \ref{fig3}(c) and \ref{fig3}(g).  We emphasize that similar type-II nodal lines have also been reported in AA stacked graphite, the high-temperature superconductor MgB$_2$ and its iso-structural counterparts such as AlB$_2$. \citep{AA_graphite,MgB2_nodalline,AlB2_nodalline} 

We present the Fermi surface of Pt$_2$HgSe$_3$ in Fig. \ref{fig3}(d) with unique electron and hole pockets that originate from both NL$_{C}$ and NL$_{\rm KH}$ nodal lines. 
{Such a Fermi surface may lead to balanced electron-hole resonance conditions and it could thus induce unusual transport signatures such as a large positive unsaturated magnetoresistance. }

Figures~\ref{fig3}(f) and~\ref{fig3}(h) show the energy bands with SOC along the selected $\kb$ paths of  NL$_{C}$ and NL$_{\rm KH}$, respectively. Clearly, the SOC opens an energy gap at the nodal crossing points. This bandgap opening facilitates the calculation of symmetry-based indicators (SI) to determine the topological state of the system. Following Ref.[\!\!\!\citenum{Chen_Fang_SI}], the band insulators in space group $P\bar{3}m1$  are defined by three $\mathcal{Z}_2$ and a single $\mathcal{Z}_4$ indicator {\it i.e.} ($\mathcal{Z}_2,\mathcal{Z}_2,\mathcal{Z}_2,\mathcal{Z}_4$). By explicitly calculating the irreducible representations of the occupied bands at different time-reversal invariant momentum points, we find ($\mathcal{Z}_2,\mathcal{Z}_2,\mathcal{Z}_2,\mathcal{Z}_4)=(0,0,1,2)$.\cite{Catalouge_TM,TQC_website,Viswanath_database} Such an SI leads to two distinct scenarios for the existence of a dual topological phase characterized by weak invariants along with {either a nonzero mirror Chern number $n_{M}=\pm2$, or a nonzero rotation invariant, $n_{2_{100}}=1$.\cite{Chen_Fang_SI} In both cases, the inversion invariant has a non-zero value ($n_i=1$).  In order to pin down the exact topological state, we further calculated the mirror Chern number, $n_M$, and found it to be $-2$. The calculated SI and topological invariants are listed in Table \ref{T1}. Thus, the topological phase of Pt$_2$HgSe$_3$ is characterized by both (001) weak topological invariants and a non-zero mirror Chern number {$n_M=-2$}. }

\begin{table}[t]
\caption{Calculated symmetry indicator  and topological invariants for Pt$_2$HgSe$_3$.} 
\begin{tabular}{c c c c c c}
\hline \hline 
 $(\mathcal{Z}_2,\mathcal{Z}_2,\mathcal{Z}_2,\mathcal{Z}_4)$ & $(\nu_0;\nu_1,\nu_2,\nu_3)$ & $n_{M}$ & $n_{2_{100}}$ & $n_i$ \\ 
\hline 
(0,0,1,2) & (0;001) & {$-2$} & 0 & 1 \\ 
\hline  \hline
\end{tabular} 
\label{T1}
\end{table}

\section{Surface electronic structure} \label{TSSs}

\begin{figure}[ht!]
\includegraphics[width=1.0\linewidth]{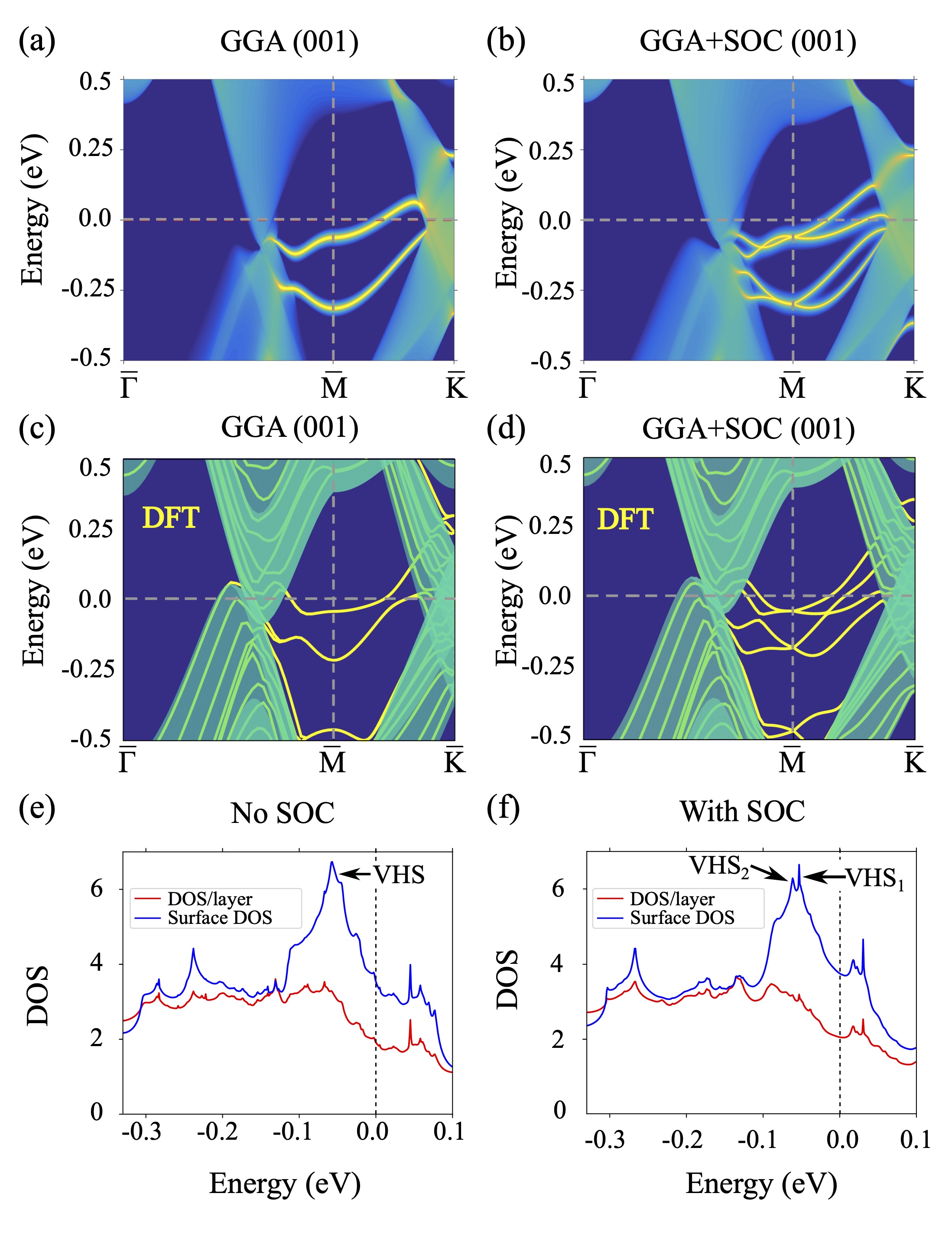}
\caption{Surface band structure of the (001) semi-infinite slab calculated using the Green's function method (a) without and (b) with SOC. Sharp yellow lines represent surface states. {The projected bulk NL$_C$ nodal line crossing is resolved along $\bar{\Gamma}\bar{M}$ in (a).} Surface band structure of a 10L thick Pt$_2$HgSe$_3$ slab obtained via first-principles calculations (c) without and (d) with SOC. The shaded-green region highlights the projected bulk bands and solid-yellow lines mark the surface states. 
{ Density-of-states (DOS) for the 10-layer slab (e) without and (f) with SOC. Average DOS per layer is shown along with the DOS from only the surface layer. The contribution of the VHSs associated with the saddle-points in the surface states is marked.}}
\label{fig4}
\end{figure}

We present the electronic spectrum of the {(001) surface of Pt$_2$HgSe$_3$ in Fig.~\ref{fig4} with Hg surface termination, which is the natural cleavage plane (Fig. \ref{fig1})}.  The projection of the NL$_{\rm C}$ nodal lines on the (001) surface forms two closed loops whereas the NL$_{\rm KH}$ nodal lines project at the corner points of the (001) surface BZ. Topological surface states related to NL$_{\rm C}$ are therefore more obvious over the (001) surface as seen in Fig.~\ref{fig4}(a) (without SOC). The two drumhead surface states (DSSs) nested outside the nodal lines are clearly visible, consistent with the calculated nontrivial character of the nodal lines. {Interestingly, the DSS which lies closer to the Fermi energy has opposite band curvatures along the $\overline{M}-\overline{\Gamma}$ and $\overline{M}-\overline{K}$ directions. Specifically, this DSS has a maximum at the $\overline{M}$ point if one looks from the $\overline{\Gamma}-\overline{M}$ direction whereas the $\overline{M}$ point is a minimum when approached from the $\overline{K}-\overline{M}$ direction. The DSS thus forms a unique saddle-like $E_{\bf k}$ dispersion around the $\overline{M}$ point.} When the SOC is included in the computations, the DSSs split away from the $\T$-symmetric $\overline{M}$ point [see Fig. \ref{fig4}(b)], and evolve into topological states with saddle-like energy dispersion, see Sec. V for more details. {The existence of two Dirac-like surface-state crossings at the $\overline{M}$ point is in accord with the bulk nonzero mirror Chern number of $c_M = -2$.}

Tight-binding based methods for calculating surface spectrum generally neglect effects on the surface potential due to charge redistribution near the surface. In order to emphasize the robustness of our saddle-like topological states, we calculated the electronic structure of a 10-layer slab on a first-principles basis, where effects of surface charge redistribution are included self-consistently. Figs.~\ref{fig4}(c) and~\ref{fig4}(d) show reasonable agreement between the results of tight-binding and first-principles computations, at least insofar as the saddle-like energy dispersion of the DSSs is concerned. 
{Figs. \ref{fig4}(e) and \ref{fig4}(f) show the calculated DOS for the 10-layer slab without and with the SOC, respectively. The high DOS near the saddle-points reflects the presence of the VHSs. The single VHS feature in the DOS (without SOC) splits into two VHSs in Fig. \ref{fig4}(e) after the SOC is included due to the appearance of more saddle-points in the underlying energy spectrum, see Sec. \ref{modelTSSs} for details. These results demonstrate that the surface-state VHSs yield significant features in the total DOS.}
The saddle-like surface states and the associated VHSs would, therefore, be accessible in spectroscopic experiments.

\begin{figure}[ht!]
\includegraphics[width=1.0\linewidth]{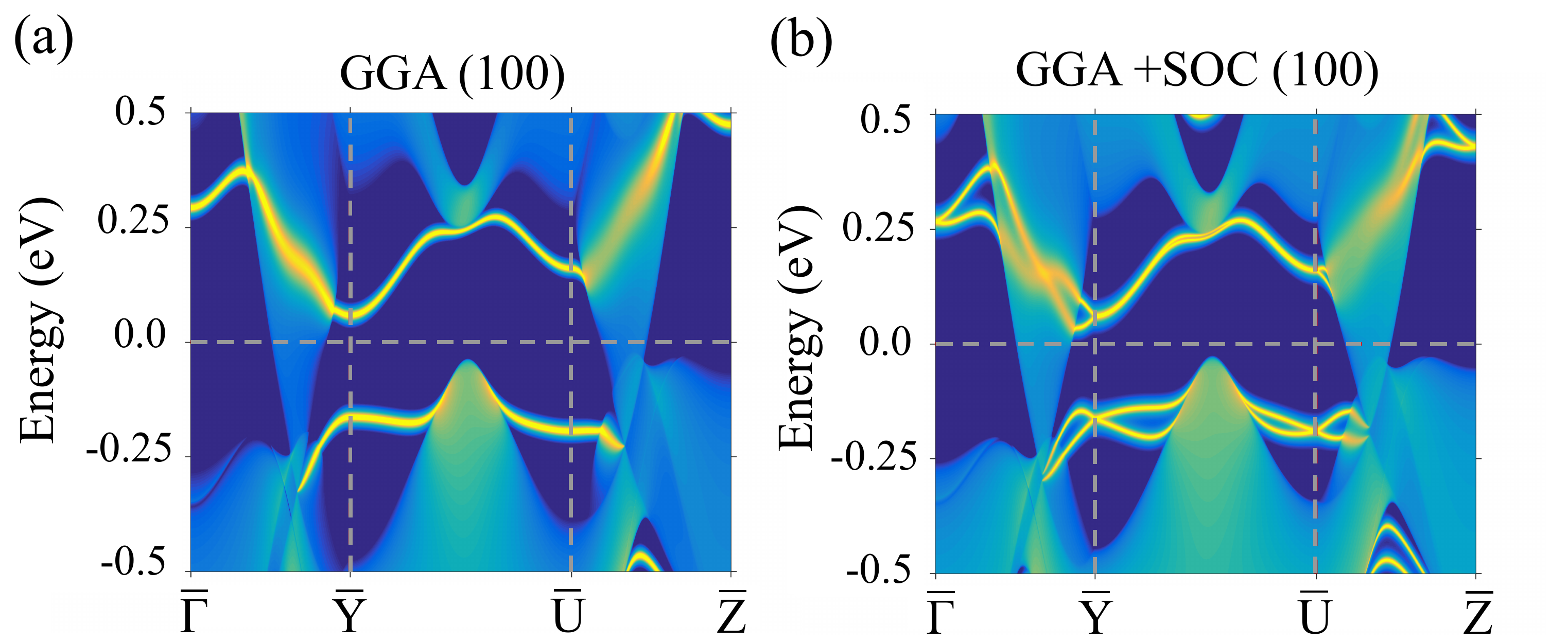}
\caption{ Surface band structure of the (100) semi-infinite slab calculated using the Green's function method (a) without and (b) with SOC. The drumhead surface states connecting the bulk NL$_{\rm KH}$ nodal line projections inside the valence band region are visible. When the SOC is included, these states evolve into topological Dirac cones at the $\overline{Y}$ and $\overline{U}$ points of the (100) surface BZ.}
\label{fig42}
\end{figure}

The (100) surface band structure  {with Hg termination} is presented in Figs.~\ref{fig42}(a) and~\ref{fig42}(b) without and with the SOC effects, respectively. Over the (100) surface, the projection of NL$_{\rm KH}$ nodal lines connects $\overline{\Gamma}-\overline{Y}$ and $\overline{Z}-\overline{U}$ symmetry lines as shown in Fig.~\ref{fig3}(a). The topological DSSs connect these projections which are seen clearly in Fig.~\ref{fig42}(a). When the SOC is included, the DSSs evolve into the Dirac-cone-like states with Dirac points at the $\overline{U}$ and $\overline{Y}$ points (Fig. \ref{fig42}(b)).  {While these surface states are in accord with the weak bulk topological invariant (001), it becomes difficult to ascertain that they cross the Fermi level an odd number of times along the $\overline{\Gamma}-\overline {Y}$ or $\overline{Z}-\overline {U}$ lines as the gap in the surface spectrum closes due to the presence of projected bulk bands.\cite{z2_parity}}

\section{$\mathbf {{\bf k \cdot p}}$ model Hamiltonian} \label{modelTSSs}

We now discuss a minimal low-energy $\mathbf {k \cdot p}$ Hamiltonian for the topological surface states on the (001) surface that captures essential features of these states. Based on our first-principles calculations, the TSSs spread around the $\overline{M}$ point on the (001) surface. Therefore, a $\mathbf {k \cdot p}$ Hamiltonian around $\overline{M}=(0, \pi)$ point is sufficient to describe the TSSs. On the (001) surface at $\overline{M}$, the little group $C_s$ contain a mirror plane symmetry. In the presence of  the SOC, the symmetry operators are $\Mx = -i\tau_{0} \sigma_{1}$ and $\T = -i\sigma_{2}\K $, where $\sigma$ and $\tau$ are Pauli matrices in the spin and sublattice space, respectively. The associated symmetry-allowed basis functions for the TSSs are $\Psi ={(\psi^A_{\alpha,\uparrow}, \psi^A_{\alpha,\downarrow}, \psi^B_{\alpha,\uparrow}, \psi^B_{\alpha,\downarrow})}^T$, where A and B denote the two sublattices of the bipartite lattice. These can be expressed as 
\begin{eqnarray}\label{eq:basis}\nonumber
\begin{split}
 |\psi_{\alpha,\sigma} \rangle = \lambda_{s} | s, \sigma \rangle + \lambda_{d_{yz}} | d_{yz}, \sigma \rangle +  \lambda_{d_{x^2-y^2}} | d_{x^2-y^2}, \sigma \rangle 
\\ 
 + \lambda_{d_{z^2}} | d_{z^2}, \sigma \rangle 
\end{split}
\end{eqnarray}
Here, the subscript $s=\uparrow / \downarrow$ denotes spin-up/spin-down, respectively, and $\lambda_{s}, \lambda_{d_{yz}}, \lambda_{d_{x^2-y^2}}$, and $\lambda_{d_{z^2}} $ describe normalization coefficients. 
Using the above basis, the minimal four-band Hamiltonian around the surface Dirac point (up to second order in momentum) can be written as,
\begin{eqnarray}\label{eq:SSHamiltonian}
\begin{split}
H_{TSS}(\textbf{p}) =\ & \dfrac{1}{2m^*} (p^2_x + \eta p^2_y) + v_{R} (p_x \sigma_2 - p_y \sigma_1) \\
& + v_{33} p_x\tau_3 \sigma_{3} + \lambda_{23} p_x p_y \tau_{2}\sigma_{3} \\
& + \delta_{30} \tau_{3}\sigma_{0} + \delta_{21} \tau_{2}\sigma_{1}.
\end{split}
\end{eqnarray}
where  $\eta$, $v_R$, $v_{33}$, $\lambda_{23}$ and $\delta_{12}$ are real numbers and $v_{R}$ denotes the Rashba parameter.  {As discussed in Ref. \cite{singh18_saddle}, $\eta < 0$ ensures a saddle-like $E_k$ dispersion. Since the little group of $\overline{M}$ hosts a mirror plane without any rotational symmetry, the saddle-like energy dispersion is symmetry allowed. This however is only a necessary condition for realizing the saddle-like $E_k$ dispersion whose actual existence will depend on material properties. } The corresponding eigenenergies of $H_{TSS}(\textbf{p})$  are

\begin{figure}
\includegraphics[width=1.0\linewidth]{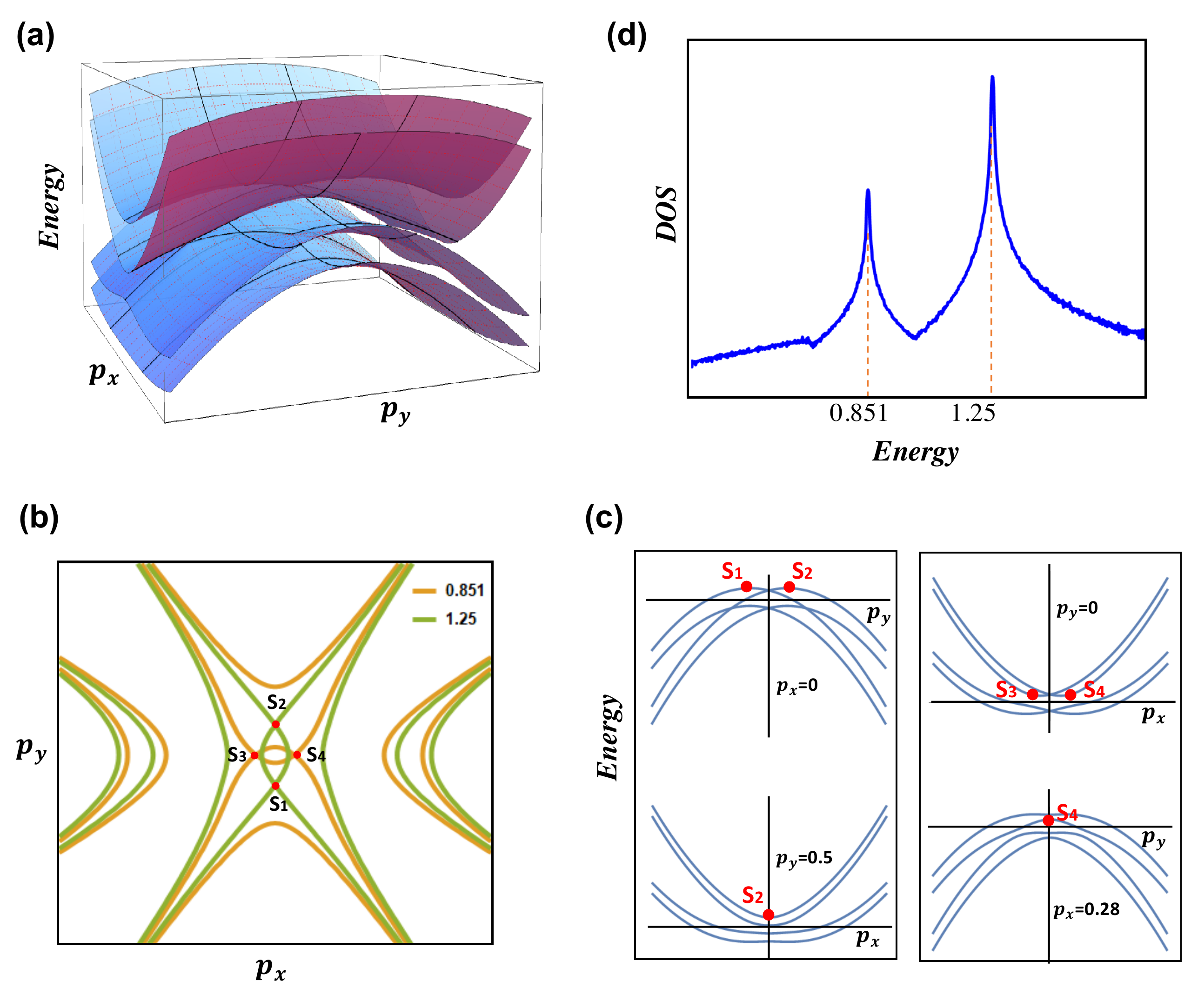}
\caption{(a) Energy dispersion of the surface state Hamiltonian $H_{TSS}(p_x, p_y)$ for {$\eta=-1, m^*=0.5,$ and $v_R= v_{33}= \lambda_{23}=\lambda_{21}=1$. These parameters, which capture the essential features of the surface states, are not obtained from first-principles calculations.} (b) The associated iso-energy contours for the upper surface states. In the presence of the SOC four saddle-points appear in the topological states. The two of these lie along the $\overline{M}-\overline{K}$ direction ($S_1$ and $S_2$) in the upper branch of the conduction band, whereas the other two belong to the lower branch and are located along the $\overline{M}-\overline{\Gamma}$ direction. {(c) Energy dispersion along $p_x$ and $p_y$ directions to emphasize the presence of saddle-points in the surface electronic spectrum.} (d) Density of states showing VHSs associated with the topological surface states.  }
\label{FigSS}
\end{figure}

\begin{widetext}
\begin{equation}\label{eq:dispersion}
E_{TSS}(\textbf{p})  = \dfrac{1}{2m^*} (p^2_x + \eta p^2_y)  + \xi \sqrt{ v^2_R p^2 +\delta^2_{21} + p^2_x (v^2_{33}+ \lambda^2_{23}p^2_y)  + \xi ' 2\delta_{21} \sqrt{v^2_R p^2_y + v^2_{33}p^2_x}}
\end{equation}
\end{widetext}
with $p^2 = p^2_x+p^2_y$ and  $\xi (\xi ') =\pm 1$. Equation~(\ref{eq:dispersion}) shows that the lower branch of the conduction band cross the top branch of the valence band at $(p_x, p_y) = (0, \pm \dfrac{\delta_{21}}{v_R})$ along the mirror invariant $\overline{M}-\overline{\Gamma}$ line. This gives rise to the Dirac cone states protected by mirror symmetry, as shown explicitly in Figs.~\ref{FigSS}(a)-\ref{FigSS}(c). In addition, for $\eta <0$, we find a pair of type II saddle-point VHSs,\cite{singh18_saddle} as illustrated in Fig. \ref{FigSS}(d) in accord with our first-principles results. {We have verified that the energy dispersions obtained by considering symmetry-allowed terms beyond the second order in the Hamiltonian retain the saddle-like features of the topological surface states with VHSs; these results are not shown in the interest of brevity.} 

\section{Topological phase transition} \label{TPTs}

\begin{figure}[t]
\includegraphics[width=1.0\linewidth]{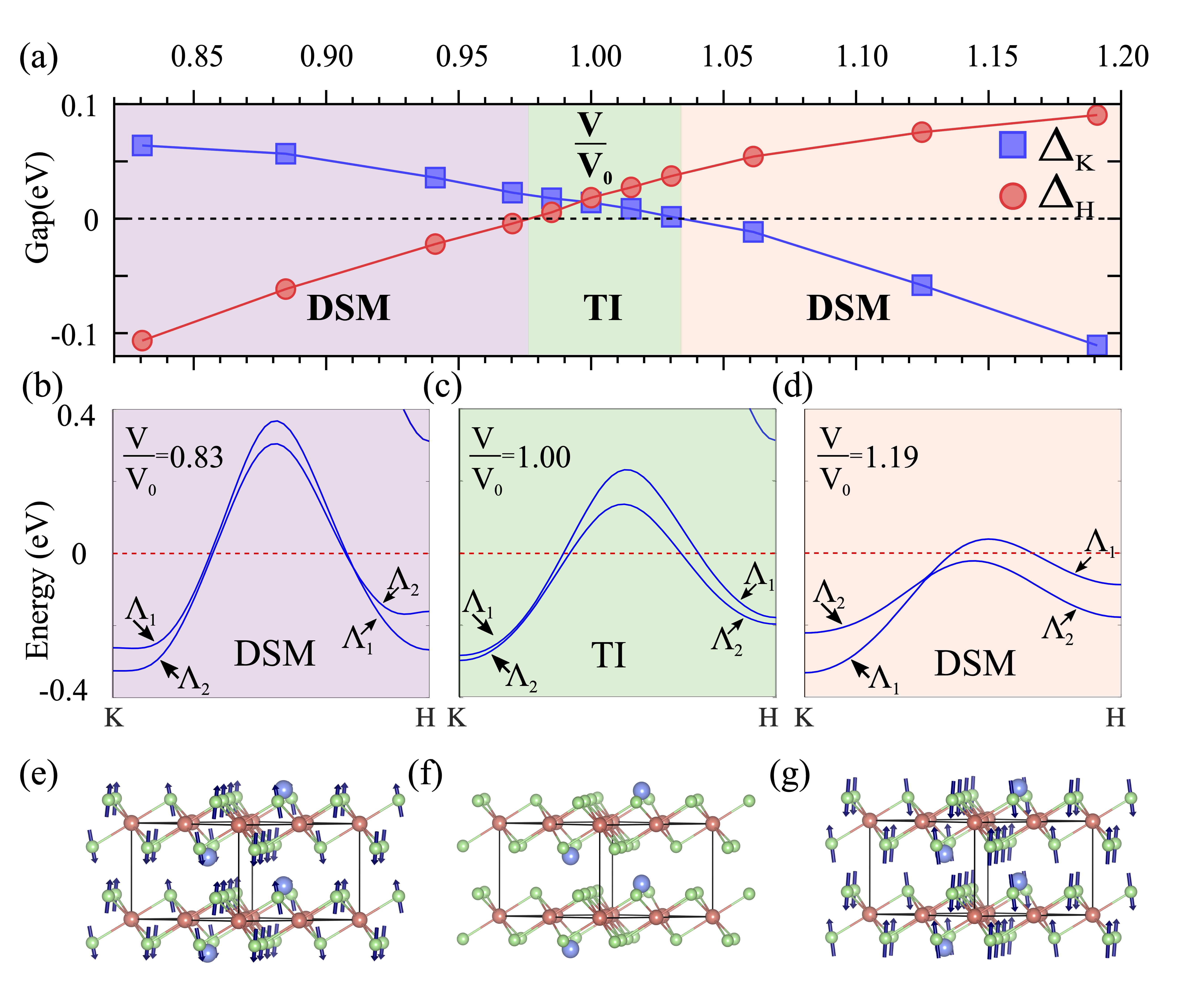}
\caption{
(a) Phase diagram showing the evolution of the bandgap $\Delta_K$ at the $K$ and $\Delta_H$ at the $H$ point with relative unit cell volume $V/V_0$. The product $\Delta_K \Delta_H$ determines the topological state. $\Delta_K \Delta_H > 0$ yields a gapped topological phase, whereas for $\Delta_K \Delta_H < 0$ we obtain a Dirac semimetal state.  Band structure along the $KH$ line for different relative cell volumes: (b) $\frac{V}{V_0}=0.83$; (c) $\frac{V}{V_0}=1.00$; and, (d) $\frac{V}{V_0}=1.19$. Panels (e)-(g) illustrate the corresponding atomic displacements with respect to the equilibrium structure. The length of the arrows is proportional to the magnitude of displacement.}
\label{fig6}
\end{figure}

We now demonstrate the possibility of tuning the topological order of Pt$_2$HgSe$_3$ and realizing a Dirac semimetal by modulating the unit cell volume with reference to Fig.~\ref{fig6}. For this purpose, it is useful to define the SOC-induced gap as $\Delta_K=E^{\Lambda_1}_K-E^{\Lambda_2}_K$ at $K$ and $\Delta_H=E^{\Lambda_1}_H-E^{\Lambda_2}_H$ at $H$ between the $\Lambda_1$ and $\Lambda_2$ states  that form a nodal line without the SOC along the $KH$ direction (see Fig.~\ref{fig6}). The evolution of $\Delta_K$ and $\Delta_H$ with relative unit cell volume $V/V_0$, where $V_0$ denotes the equilibrium unit cell volume, is presented in Fig.~\ref{fig6}(a). We find that $\Delta_K$ and $\Delta_H$ are comparable in the gapped pristine state but show opposite behavior on changing the unit cell volume ($V$). On decreasing (increasing) $V/V_0$ from its equilibrium value,  the $\Lambda_1$ and $\Lambda_2$ bands cross near $H (K) $ point and realize a tilted band crossing along the $KH$ direction, see Figs.~\ref{fig6}(b)-(d).  A detailed symmetry analysis shows that the crossing bands have opposite $C_{3z}$ rotation eigenvalues and thus the Dirac point is symmetry protected against band hybridization. Importantly, we find that when $\Delta_K \Delta_H  > 0$, the $\Lambda_1$ and $\Lambda_2$ are separated by a continuous gap and realize a gapped topological phase. But, when $\Delta_K \Delta_H  < 0$, the two states cross along the $KH$ line and the system realizes a symmetry-protected type II Dirac semimetal state. 

Our analysis suggests that $V/V_0$ can provide a control knob for continuous tuning of the position and velocity of the Dirac cones when $\Delta_K \Delta_H < 0$. Owing to the parabolic energy dispersion of $\Lambda_1$ and $\Lambda_2$, the energy location of Dirac points can be tuned to lie on the Fermi level. For $V/V_0>  0.83$, a pair of Dirac cones located on the $HKH$ line moves toward the $H$ point. These Dirac cones merge at $H$ for $V/V_0 \sim 0.975$, thereby realizing a gapped insulator state. For $V/V_0$ beyond $\sim 1.031$, the Dirac points start reappearing near the K-point \footnote{{Note that the topological phase transition to a Dirac semimetal state takes place at $\frac{V}{V_0} \sim 0.975$ ($1.031$) which corresponds to $2.5\% ~(3.1\%$) decrease (increase) of volume $V_0$. This level of strain would be practical to achieve in experiments. We have not explored the dynamical stability of the structure under pressure, although such a study will be interesting.}}.

Generally, a topological semimetal phase separates two distinct gapped insulating topological phases. In sharp contrast to this, in Pt$_2$HgSe$_3$ we find that a topological insulating phase exists as a critical region between two gapless Dirac semimetal phases. Moreover, the Dirac points in Pt$_2$HgSe$_3$ are located on BZ hinges along the $KH$ line in contrast to the other well-known Dirac semimetals such as Na$_3$Bi and PtTe$_2$ where they are located on the $\Gamma A $ line in the hexagonal BZ.\citep{Na3Bi_PRB,Cd3As2_PRB,PtTe2PRL} Such a Dirac semimetal state is unique to Pt$_2$HgSe$_3$, and has not been identified before. 

\section{Conclusion}\label{con}

In conclusion, based on our first-principles calculations combined with a ${\bf k.p}$ model Hamiltonian analysis, we identify and characterize the dual-symmetry-protected topological state of Pt$_2$HgSe$_3$. The material is shown to harbor two distinct types of nodal lines when SOC effects are neglected in the computations. Inclusion of SOC gaps out the nodal lines and drives the system into a topological phase which is characterized by both the weak topological invariant $Z_2=(0;001)$ and the mirror Chern number $n_M=-2$. The (001) surface band structure reveals the existence of unique saddle-like topological surface states with saddle-point VHSs. We also discuss the tunability of the topological state of Pt$_2$HgSe$_3$ by modulating its crystal structure. In this way, the system is shown to undergo a unique topological phase transition where a gapped topological state exists as an intermediate phase between gapless Dirac semimetal states. Our analysis suggests that the naturally cleaved (001) surface of Pt$_2$HgSe$_3$ presents an ideal testbed for exploring saddle-like topological surface states with VHSs and the associated physics in topological materials. 

 {{\it Note added:} We recently became aware of a related eprint in which dual topological state of Pt$_2$HgSe$_3$ is discussed \cite{HgPt2Se3_bulk_theory}. A more recent preprint on  
ARPES measurements of Pt$_2$HgSe$_3$ reveals the existence of saddle-like topological surface states with VHSs which are consistent with the results presented in this study \cite{HgPt2Se3_expt_ARPES}.}

\section*{ACKNOWLEDGEMENTS}
Work at the Shenzhen University is financially supported by the Shenzhen Peacock Plan (KQTD2016053112042971) and Science and Technology Planning Project of Guangdong Province (2016B050501005). The work at Northeastern University is supported by the U.S. Department of Energy (DOE), Office of Science, Basic Energy Sciences Grant No. DE-FG02-07ER46352, and benefited from Northeastern University’s Advanced Scientific Computation Center and the National Energy Research Scientific Computing Center through DOE Grant No. DE-AC02-05CH11231. T.-R.C. was supported from Young Scholar Fellowship Program by Ministry of Science and Technology (MOST) in Taiwan, under MOST Grant for the Columbus Program MOST107-2636-M-006-004, National Cheng Kung University, Taiwan, and National Center for Theoretical Sciences (NCTS), Taiwan. This work is supported partially by the MOST, Taiwan, Grants No. MOST 107-2627-E-006-001.  H. L. acknowledges Academia Sinica, Taiwan for the support under Innovative Materials and Analysis Technology Exploration (AS-iMATE-107-11). BG acknowledges the CSIR for Senior Research Fellowship. We thank CC-IITK for providing the HPC facility.\\

\appendix
\section {Structural details}\label{appA}
The experimental lattice constants and the relaxed atomic positions for bulk Pt$_2$HgSe$_3$ that have been used in our computations are listed in the table below. 

\begin{table}[h!]
\centering
\begin{ruledtabular}
\begin{tabular}{cccc}
				&   \multicolumn{2}{c}{\underline{Wyckoff positions}} \\
Atom (Wyckoff symbol) & Experiments  &  Optimized\\
 				& 	$(x,y,z)$ 		& $(x,y,z)$\\
  \hline
 Pt1 ($1a$)  &  $(0,0,0)$           & 		$(0,0,0)$ \\
 Pt2 ($3e$)  &  $(\frac{1}{2},0,0)$        &  		$(\frac{1}{2},0,0)$ \\
 Hg ($2d$)  & $(\frac{1}{3},\frac{2}{3},0.3507)$  & 		$(\frac{1}{3},\frac{2}{3},0.3526)$ \\
 Se  ($6i$)  &  $(0.8196,0.1804,0.2492)$  &  $ (0.8285,0.1716,0.2477)$ \\
\end{tabular}
\end{ruledtabular}
\label{tab:t1}
\end{table}

\bibliography{Pt2HgSe3}

\begin{thebibliography}{67}%
\makeatletter
\providecommand \@ifxundefined [1]{%
 \@ifx{#1\undefined}
}%
\providecommand \@ifnum [1]{%
 \ifnum #1\expandafter \@firstoftwo
 \else \expandafter \@secondoftwo
 \fi
}%
\providecommand \@ifx [1]{%
 \ifx #1\expandafter \@firstoftwo
 \else \expandafter \@secondoftwo
 \fi
}%
\providecommand \natexlab [1]{#1}%
\providecommand \enquote  [1]{``#1''}%
\providecommand \bibnamefont  [1]{#1}%
\providecommand \bibfnamefont [1]{#1}%
\providecommand \citenamefont [1]{#1}%
\providecommand \href@noop [0]{\@secondoftwo}%
\providecommand \href [0]{\begingroup \@sanitize@url \@href}%
\providecommand \@href[1]{\@@startlink{#1}\@@href}%
\providecommand \@@href[1]{\endgroup#1\@@endlink}%
\providecommand \@sanitize@url [0]{\catcode `\\12\catcode `\$12\catcode
  `\&12\catcode `\#12\catcode `\^12\catcode `\_12\catcode `\%12\relax}%
\providecommand \@@startlink[1]{}%
\providecommand \@@endlink[0]{}%
\providecommand \url  [0]{\begingroup\@sanitize@url \@url }%
\providecommand \@url [1]{\endgroup\@href {#1}{\urlprefix }}%
\providecommand \urlprefix  [0]{URL }%
\providecommand \Eprint [0]{\href }%
\providecommand \doibase [0]{http://dx.doi.org/}%
\providecommand \selectlanguage [0]{\@gobble}%
\providecommand \bibinfo  [0]{\@secondoftwo}%
\providecommand \bibfield  [0]{\@secondoftwo}%
\providecommand \translation [1]{[#1]}%
\providecommand \BibitemOpen [0]{}%
\providecommand \bibitemStop [0]{}%
\providecommand \bibitemNoStop [0]{.\EOS\space}%
\providecommand \EOS [0]{\spacefactor3000\relax}%
\providecommand \BibitemShut  [1]{\csname bibitem#1\endcsname}%
\let\auto@bib@innerbib\@empty
\bibitem [{\citenamefont {Bansil}\ \emph {et~al.}(2016)\citenamefont {Bansil},
  \citenamefont {Lin},\ and\ \citenamefont {Das}}]{Bansil16}%
  \BibitemOpen
  \bibfield  {author} {\bibinfo {author} {\bibfnamefont {A.}~\bibnamefont
  {Bansil}}, \bibinfo {author} {\bibfnamefont {Hsin}\ \bibnamefont {Lin}}, \
  and\ \bibinfo {author} {\bibfnamefont {Tanmoy}\ \bibnamefont {Das}},\
  }\bibfield  {title} {\enquote {\bibinfo {title} {Colloquium},}\ }\href
  {\doibase 10.1103/RevModPhys.88.021004} {\bibfield  {journal} {\bibinfo
  {journal} {Rev. Mod. Phys.}\ }\textbf {\bibinfo {volume} {88}},\ \bibinfo
  {pages} {021004} (\bibinfo {year} {2016})}\BibitemShut {NoStop}%
\bibitem [{\citenamefont {Hasan}\ and\ \citenamefont {Kane}(2010)}]{Hasan10}%
  \BibitemOpen
  \bibfield  {author} {\bibinfo {author} {\bibfnamefont {M.~Z.}\ \bibnamefont
  {Hasan}}\ and\ \bibinfo {author} {\bibfnamefont {C.~L.}\ \bibnamefont
  {Kane}},\ }\bibfield  {title} {\enquote {\bibinfo {title} {Colloquium},}\
  }\href {\doibase 10.1103/RevModPhys.82.3045} {\bibfield  {journal} {\bibinfo
  {journal} {Rev. Mod. Phys.}\ }\textbf {\bibinfo {volume} {82}},\ \bibinfo
  {pages} {3045--3067} (\bibinfo {year} {2010})}\BibitemShut {NoStop}%
\bibitem [{\citenamefont {Qi}\ and\ \citenamefont {Zhang}(2011)}]{Qi11}%
  \BibitemOpen
  \bibfield  {author} {\bibinfo {author} {\bibfnamefont {Xiao-Liang}\
  \bibnamefont {Qi}}\ and\ \bibinfo {author} {\bibfnamefont {Shou-Cheng}\
  \bibnamefont {Zhang}},\ }\bibfield  {title} {\enquote {\bibinfo {title}
  {Topological insulators and superconductors},}\ }\href {\doibase
  10.1103/RevModPhys.83.1057} {\bibfield  {journal} {\bibinfo  {journal} {Rev.
  Mod. Phys.}\ }\textbf {\bibinfo {volume} {83}},\ \bibinfo {pages}
  {1057--1110} (\bibinfo {year} {2011})}\BibitemShut {NoStop}%
\bibitem [{\citenamefont {Fu}\ \emph {et~al.}(2007)\citenamefont {Fu},
  \citenamefont {Kane},\ and\ \citenamefont {Mele}}]{Fu07}%
  \BibitemOpen
  \bibfield  {author} {\bibinfo {author} {\bibfnamefont {Liang}\ \bibnamefont
  {Fu}}, \bibinfo {author} {\bibfnamefont {C.~L.}\ \bibnamefont {Kane}}, \ and\
  \bibinfo {author} {\bibfnamefont {E.~J.}\ \bibnamefont {Mele}},\ }\bibfield
  {title} {\enquote {\bibinfo {title} {Topological insulators in three
  dimensions},}\ }\href {\doibase 10.1103/PhysRevLett.98.106803} {\bibfield
  {journal} {\bibinfo  {journal} {Phys. Rev. Lett.}\ }\textbf {\bibinfo
  {volume} {98}},\ \bibinfo {pages} {106803} (\bibinfo {year}
  {2007})}\BibitemShut {NoStop}%
\bibitem [{\citenamefont {Fu}(2011)}]{Fu_TCI}%
  \BibitemOpen
  \bibfield  {author} {\bibinfo {author} {\bibfnamefont {Liang}\ \bibnamefont
  {Fu}},\ }\bibfield  {title} {\enquote {\bibinfo {title} {Topological
  crystalline insulators},}\ }\href {\doibase 10.1103/PhysRevLett.106.106802}
  {\bibfield  {journal} {\bibinfo  {journal} {Phys. Rev. Lett.}\ }\textbf
  {\bibinfo {volume} {106}},\ \bibinfo {pages} {106802} (\bibinfo {year}
  {2011})}\BibitemShut {NoStop}%
\bibitem [{\citenamefont {Po}\ \emph {et~al.}(2017)\citenamefont {Po},
  \citenamefont {Vishwanath},\ and\ \citenamefont {Watanabe}}]{Po2017}%
  \BibitemOpen
  \bibfield  {author} {\bibinfo {author} {\bibfnamefont {Hoi~Chun}\
  \bibnamefont {Po}}, \bibinfo {author} {\bibfnamefont {Ashvin}\ \bibnamefont
  {Vishwanath}}, \ and\ \bibinfo {author} {\bibfnamefont {Haruki}\ \bibnamefont
  {Watanabe}},\ }\bibfield  {title} {\enquote {\bibinfo {title} {Symmetry-based
  indicators of band topology in the 230 space groups},}\ }\href {\doibase
  10.1038/s41467-017-00133-2} {\bibfield  {journal} {\bibinfo  {journal} {Nat.
  Commun.}\ }\textbf {\bibinfo {volume} {8}},\ \bibinfo {pages} {50} (\bibinfo
  {year} {2017})}\BibitemShut {NoStop}%
\bibitem [{\citenamefont {Song}\ \emph {et~al.}(2018)\citenamefont {Song},
  \citenamefont {Zhang}, \citenamefont {Fang},\ and\ \citenamefont
  {Fang}}]{Chen_Fang_SI}%
  \BibitemOpen
  \bibfield  {author} {\bibinfo {author} {\bibfnamefont {Zhida}\ \bibnamefont
  {Song}}, \bibinfo {author} {\bibfnamefont {Tiantian}\ \bibnamefont {Zhang}},
  \bibinfo {author} {\bibfnamefont {Zhong}\ \bibnamefont {Fang}}, \ and\
  \bibinfo {author} {\bibfnamefont {Chen}\ \bibnamefont {Fang}},\ }\bibfield
  {title} {\enquote {\bibinfo {title} {Quantitative mappings between symmetry
  and topology in solids},}\ }\href {\doibase 10.1038/s41467-018-06010-w}
  {\bibfield  {journal} {\bibinfo  {journal} {Nat. Commun.}\ }\textbf {\bibinfo
  {volume} {9}},\ \bibinfo {pages} {3530} (\bibinfo {year} {2018})}\BibitemShut
  {NoStop}%
\bibitem [{\citenamefont {Kruthoff}\ \emph {et~al.}(2017)\citenamefont
  {Kruthoff}, \citenamefont {de~Boer}, \citenamefont {van Wezel}, \citenamefont
  {Kane},\ and\ \citenamefont {Slager}}]{PhysRevX.7.041069}%
  \BibitemOpen
  \bibfield  {author} {\bibinfo {author} {\bibfnamefont {Jorrit}\ \bibnamefont
  {Kruthoff}}, \bibinfo {author} {\bibfnamefont {Jan}\ \bibnamefont {de~Boer}},
  \bibinfo {author} {\bibfnamefont {Jasper}\ \bibnamefont {van Wezel}},
  \bibinfo {author} {\bibfnamefont {Charles~L.}\ \bibnamefont {Kane}}, \ and\
  \bibinfo {author} {\bibfnamefont {Robert-Jan}\ \bibnamefont {Slager}},\
  }\bibfield  {title} {\enquote {\bibinfo {title} {Topological classification
  of crystalline insulators through band structure combinatorics},}\ }\href
  {\doibase 10.1103/PhysRevX.7.041069} {\bibfield  {journal} {\bibinfo
  {journal} {Phys. Rev. X}\ }\textbf {\bibinfo {volume} {7}},\ \bibinfo {pages}
  {041069} (\bibinfo {year} {2017})}\BibitemShut {NoStop}%
\bibitem [{\citenamefont {Hsieh}\ \emph {et~al.}(2012)\citenamefont {Hsieh},
  \citenamefont {Lin}, \citenamefont {Liu}, \citenamefont {Duan}, \citenamefont
  {Bansil},\ and\ \citenamefont {Fu}}]{Hsieh2012}%
  \BibitemOpen
  \bibfield  {author} {\bibinfo {author} {\bibfnamefont {Timothy~H.}\
  \bibnamefont {Hsieh}}, \bibinfo {author} {\bibfnamefont {Hsin}\ \bibnamefont
  {Lin}}, \bibinfo {author} {\bibfnamefont {Junwei}\ \bibnamefont {Liu}},
  \bibinfo {author} {\bibfnamefont {Wenhui}\ \bibnamefont {Duan}}, \bibinfo
  {author} {\bibfnamefont {Arun}\ \bibnamefont {Bansil}}, \ and\ \bibinfo
  {author} {\bibfnamefont {Liang}\ \bibnamefont {Fu}},\ }\bibfield  {title}
  {\enquote {\bibinfo {title} {Topological crystalline insulators in the snte
  material class},}\ }\href {https://doi.org/10.1038/ncomms1969} {\bibfield
  {journal} {\bibinfo  {journal} {Nat. Commun.}\ }\textbf {\bibinfo {volume}
  {3}},\ \bibinfo {pages} {982} (\bibinfo {year} {2012})}\BibitemShut {NoStop}%
\bibitem [{\citenamefont {Young}\ \emph {et~al.}(2012)\citenamefont {Young},
  \citenamefont {Zaheer}, \citenamefont {Teo}, \citenamefont {Kane},
  \citenamefont {Mele},\ and\ \citenamefont {Rappe}}]{Kane_DSM_3D}%
  \BibitemOpen
  \bibfield  {author} {\bibinfo {author} {\bibfnamefont {S.~M.}\ \bibnamefont
  {Young}}, \bibinfo {author} {\bibfnamefont {S.}~\bibnamefont {Zaheer}},
  \bibinfo {author} {\bibfnamefont {J.~C.~Y.}\ \bibnamefont {Teo}}, \bibinfo
  {author} {\bibfnamefont {C.~L.}\ \bibnamefont {Kane}}, \bibinfo {author}
  {\bibfnamefont {E.~J.}\ \bibnamefont {Mele}}, \ and\ \bibinfo {author}
  {\bibfnamefont {A.~M.}\ \bibnamefont {Rappe}},\ }\bibfield  {title} {\enquote
  {\bibinfo {title} {Dirac semimetal in three dimensions},}\ }\href {\doibase
  10.1103/PhysRevLett.108.140405} {\bibfield  {journal} {\bibinfo  {journal}
  {Phys. Rev. Lett.}\ }\textbf {\bibinfo {volume} {108}},\ \bibinfo {pages}
  {140405} (\bibinfo {year} {2012})}\BibitemShut {NoStop}%
\bibitem [{\citenamefont {Armitage}\ \emph {et~al.}(2018)\citenamefont
  {Armitage}, \citenamefont {Mele},\ and\ \citenamefont
  {Vishwanath}}]{DSM_WSM_review_viswanath}%
  \BibitemOpen
  \bibfield  {author} {\bibinfo {author} {\bibfnamefont {N.~P.}\ \bibnamefont
  {Armitage}}, \bibinfo {author} {\bibfnamefont {E.~J.}\ \bibnamefont {Mele}},
  \ and\ \bibinfo {author} {\bibfnamefont {Ashvin}\ \bibnamefont
  {Vishwanath}},\ }\bibfield  {title} {\enquote {\bibinfo {title} {Weyl and
  dirac semimetals in three-dimensional solids},}\ }\href {\doibase
  10.1103/RevModPhys.90.015001} {\bibfield  {journal} {\bibinfo  {journal}
  {Rev. Mod. Phys.}\ }\textbf {\bibinfo {volume} {90}},\ \bibinfo {pages}
  {015001} (\bibinfo {year} {2018})}\BibitemShut {NoStop}%
\bibitem [{\citenamefont {Wang}\ \emph {et~al.}(2012)\citenamefont {Wang},
  \citenamefont {Sun}, \citenamefont {Chen}, \citenamefont {Franchini},
  \citenamefont {Xu}, \citenamefont {Weng}, \citenamefont {Dai},\ and\
  \citenamefont {Fang}}]{Na3Bi_PRB}%
  \BibitemOpen
  \bibfield  {author} {\bibinfo {author} {\bibfnamefont {Zhijun}\ \bibnamefont
  {Wang}}, \bibinfo {author} {\bibfnamefont {Yan}\ \bibnamefont {Sun}},
  \bibinfo {author} {\bibfnamefont {Xing-Qiu}\ \bibnamefont {Chen}}, \bibinfo
  {author} {\bibfnamefont {Cesare}\ \bibnamefont {Franchini}}, \bibinfo
  {author} {\bibfnamefont {Gang}\ \bibnamefont {Xu}}, \bibinfo {author}
  {\bibfnamefont {Hongming}\ \bibnamefont {Weng}}, \bibinfo {author}
  {\bibfnamefont {Xi}~\bibnamefont {Dai}}, \ and\ \bibinfo {author}
  {\bibfnamefont {Zhong}\ \bibnamefont {Fang}},\ }\bibfield  {title} {\enquote
  {\bibinfo {title} {Dirac semimetal and topological phase transitions in
  ${A}_{3}$bi ($a=\text{Na}$, k, rb)},}\ }\href {\doibase
  10.1103/PhysRevB.85.195320} {\bibfield  {journal} {\bibinfo  {journal} {Phys.
  Rev. B}\ }\textbf {\bibinfo {volume} {85}},\ \bibinfo {pages} {195320}
  (\bibinfo {year} {2012})}\BibitemShut {NoStop}%
\bibitem [{\citenamefont {Wang}\ \emph {et~al.}(2013)\citenamefont {Wang},
  \citenamefont {Weng}, \citenamefont {Wu}, \citenamefont {Dai},\ and\
  \citenamefont {Fang}}]{Cd3As2_PRB}%
  \BibitemOpen
  \bibfield  {author} {\bibinfo {author} {\bibfnamefont {Zhijun}\ \bibnamefont
  {Wang}}, \bibinfo {author} {\bibfnamefont {Hongming}\ \bibnamefont {Weng}},
  \bibinfo {author} {\bibfnamefont {Quansheng}\ \bibnamefont {Wu}}, \bibinfo
  {author} {\bibfnamefont {Xi}~\bibnamefont {Dai}}, \ and\ \bibinfo {author}
  {\bibfnamefont {Zhong}\ \bibnamefont {Fang}},\ }\bibfield  {title} {\enquote
  {\bibinfo {title} {Three-dimensional dirac semimetal and quantum transport in
  cd${}_{3}$as${}_{2}$},}\ }\href {\doibase 10.1103/PhysRevB.88.125427}
  {\bibfield  {journal} {\bibinfo  {journal} {Phys. Rev. B}\ }\textbf {\bibinfo
  {volume} {88}},\ \bibinfo {pages} {125427} (\bibinfo {year}
  {2013})}\BibitemShut {NoStop}%
\bibitem [{\citenamefont {Singh}\ \emph {et~al.}(2012)\citenamefont {Singh},
  \citenamefont {Sharma}, \citenamefont {Lin}, \citenamefont {Hasan},
  \citenamefont {Prasad},\ and\ \citenamefont {Bansil}}]{singh12}%
  \BibitemOpen
  \bibfield  {author} {\bibinfo {author} {\bibfnamefont {Bahadur}\ \bibnamefont
  {Singh}}, \bibinfo {author} {\bibfnamefont {Ashutosh}\ \bibnamefont
  {Sharma}}, \bibinfo {author} {\bibfnamefont {H.}~\bibnamefont {Lin}},
  \bibinfo {author} {\bibfnamefont {M.~Z.}\ \bibnamefont {Hasan}}, \bibinfo
  {author} {\bibfnamefont {R.}~\bibnamefont {Prasad}}, \ and\ \bibinfo {author}
  {\bibfnamefont {A.}~\bibnamefont {Bansil}},\ }\bibfield  {title} {\enquote
  {\bibinfo {title} {Topological electronic structure and weyl semimetal in the
  tlbise${}_{2}$ class of semiconductors},}\ }\href {\doibase
  10.1103/PhysRevB.86.115208} {\bibfield  {journal} {\bibinfo  {journal} {Phys.
  Rev. B}\ }\textbf {\bibinfo {volume} {86}},\ \bibinfo {pages} {115208}
  (\bibinfo {year} {2012})}\BibitemShut {NoStop}%
\bibitem [{\citenamefont {Fang}\ \emph {et~al.}(2015)\citenamefont {Fang},
  \citenamefont {Chen}, \citenamefont {Kee},\ and\ \citenamefont
  {Fu}}]{Z2NL_Fu}%
  \BibitemOpen
  \bibfield  {author} {\bibinfo {author} {\bibfnamefont {Chen}\ \bibnamefont
  {Fang}}, \bibinfo {author} {\bibfnamefont {Yige}\ \bibnamefont {Chen}},
  \bibinfo {author} {\bibfnamefont {Hae-Young}\ \bibnamefont {Kee}}, \ and\
  \bibinfo {author} {\bibfnamefont {Liang}\ \bibnamefont {Fu}},\ }\bibfield
  {title} {\enquote {\bibinfo {title} {Topological nodal line semimetals with
  and without spin-orbital coupling},}\ }\href {\doibase
  10.1103/PhysRevB.92.081201} {\bibfield  {journal} {\bibinfo  {journal} {Phys.
  Rev. B}\ }\textbf {\bibinfo {volume} {92}},\ \bibinfo {pages} {081201}
  (\bibinfo {year} {2015})}\BibitemShut {NoStop}%
\bibitem [{\citenamefont {Kim}\ \emph {et~al.}(2015)\citenamefont {Kim},
  \citenamefont {Wieder}, \citenamefont {Kane},\ and\ \citenamefont
  {Rappe}}]{Z2NL_Kane}%
  \BibitemOpen
  \bibfield  {author} {\bibinfo {author} {\bibfnamefont {Youngkuk}\
  \bibnamefont {Kim}}, \bibinfo {author} {\bibfnamefont {Benjamin~J.}\
  \bibnamefont {Wieder}}, \bibinfo {author} {\bibfnamefont {C.~L.}\
  \bibnamefont {Kane}}, \ and\ \bibinfo {author} {\bibfnamefont {Andrew~M.}\
  \bibnamefont {Rappe}},\ }\bibfield  {title} {\enquote {\bibinfo {title}
  {Dirac line nodes in inversion-symmetric crystals},}\ }\href {\doibase
  10.1103/PhysRevLett.115.036806} {\bibfield  {journal} {\bibinfo  {journal}
  {Phys. Rev. Lett.}\ }\textbf {\bibinfo {volume} {115}},\ \bibinfo {pages}
  {036806} (\bibinfo {year} {2015})}\BibitemShut {NoStop}%
\bibitem [{\citenamefont {Fang}\ \emph {et~al.}(2016)\citenamefont {Fang},
  \citenamefont {Weng}, \citenamefont {Dai},\ and\ \citenamefont
  {Fang}}]{Fang_2016}%
  \BibitemOpen
  \bibfield  {author} {\bibinfo {author} {\bibfnamefont {Chen}\ \bibnamefont
  {Fang}}, \bibinfo {author} {\bibfnamefont {Hongming}\ \bibnamefont {Weng}},
  \bibinfo {author} {\bibfnamefont {Xi}~\bibnamefont {Dai}}, \ and\ \bibinfo
  {author} {\bibfnamefont {Zhong}\ \bibnamefont {Fang}},\ }\bibfield  {title}
  {\enquote {\bibinfo {title} {Topological nodal line semimetals},}\ }\href
  {\doibase 10.1088/1674-1056/25/11/117106} {\bibfield  {journal} {\bibinfo
  {journal} {Chinese Physics B}\ }\textbf {\bibinfo {volume} {25}},\ \bibinfo
  {pages} {117106} (\bibinfo {year} {2016})}\BibitemShut {NoStop}%
\bibitem [{\citenamefont {Wang}\ \emph {et~al.}(2017)\citenamefont {Wang},
  \citenamefont {Jian},\ and\ \citenamefont {Yao}}]{Wang17}%
  \BibitemOpen
  \bibfield  {author} {\bibinfo {author} {\bibfnamefont {Luyang}\ \bibnamefont
  {Wang}}, \bibinfo {author} {\bibfnamefont {Shao-Kai}\ \bibnamefont {Jian}}, \
  and\ \bibinfo {author} {\bibfnamefont {Hong}\ \bibnamefont {Yao}},\
  }\bibfield  {title} {\enquote {\bibinfo {title} {Hourglass semimetals with
  nonsymmorphic symmetries in three dimensions},}\ }\href {\doibase
  10.1103/PhysRevB.96.075110} {\bibfield  {journal} {\bibinfo  {journal} {Phys.
  Rev. B}\ }\textbf {\bibinfo {volume} {96}},\ \bibinfo {pages} {075110}
  (\bibinfo {year} {2017})}\BibitemShut {NoStop}%
\bibitem [{\citenamefont {Singh}\ \emph
  {et~al.}(2018{\natexlab{a}})\citenamefont {Singh}, \citenamefont {Ghosh},
  \citenamefont {Su}, \citenamefont {Lin}, \citenamefont {Agarwal},\ and\
  \citenamefont {Bansil}}]{hourglass_Ag2BiO3}%
  \BibitemOpen
  \bibfield  {author} {\bibinfo {author} {\bibfnamefont {Bahadur}\ \bibnamefont
  {Singh}}, \bibinfo {author} {\bibfnamefont {Barun}\ \bibnamefont {Ghosh}},
  \bibinfo {author} {\bibfnamefont {Chenliang}\ \bibnamefont {Su}}, \bibinfo
  {author} {\bibfnamefont {Hsin}\ \bibnamefont {Lin}}, \bibinfo {author}
  {\bibfnamefont {Amit}\ \bibnamefont {Agarwal}}, \ and\ \bibinfo {author}
  {\bibfnamefont {Arun}\ \bibnamefont {Bansil}},\ }\bibfield  {title} {\enquote
  {\bibinfo {title} {Topological hourglass dirac semimetal in the nonpolar
  phase of ${\mathrm{ag}}_{2}{\mathrm{bio}}_{3}$},}\ }\href {\doibase
  10.1103/PhysRevLett.121.226401} {\bibfield  {journal} {\bibinfo  {journal}
  {Phys. Rev. Lett.}\ }\textbf {\bibinfo {volume} {121}},\ \bibinfo {pages}
  {226401} (\bibinfo {year} {2018}{\natexlab{a}})}\BibitemShut {NoStop}%
\bibitem [{\citenamefont {Zhu}\ \emph {et~al.}(2016)\citenamefont {Zhu},
  \citenamefont {Winkler}, \citenamefont {Wu}, \citenamefont {Li},\ and\
  \citenamefont {Soluyanov}}]{triplepoint_1}%
  \BibitemOpen
  \bibfield  {author} {\bibinfo {author} {\bibfnamefont {Ziming}\ \bibnamefont
  {Zhu}}, \bibinfo {author} {\bibfnamefont {Georg~W.}\ \bibnamefont {Winkler}},
  \bibinfo {author} {\bibfnamefont {QuanSheng}\ \bibnamefont {Wu}}, \bibinfo
  {author} {\bibfnamefont {Ju}~\bibnamefont {Li}}, \ and\ \bibinfo {author}
  {\bibfnamefont {Alexey~A.}\ \bibnamefont {Soluyanov}},\ }\bibfield  {title}
  {\enquote {\bibinfo {title} {Triple point topological metals},}\ }\href
  {\doibase 10.1103/PhysRevX.6.031003} {\bibfield  {journal} {\bibinfo
  {journal} {Phys. Rev. X}\ }\textbf {\bibinfo {volume} {6}},\ \bibinfo {pages}
  {031003} (\bibinfo {year} {2016})}\BibitemShut {NoStop}%
\bibitem [{\citenamefont {Tanaka}\ \emph {et~al.}(2012)\citenamefont {Tanaka},
  \citenamefont {Ren}, \citenamefont {Sato}, \citenamefont {Nakayama},
  \citenamefont {Souma}, \citenamefont {Takahashi}, \citenamefont {Segawa},\
  and\ \citenamefont {Ando}}]{Tanaka2012}%
  \BibitemOpen
  \bibfield  {author} {\bibinfo {author} {\bibfnamefont {Y.}~\bibnamefont
  {Tanaka}}, \bibinfo {author} {\bibfnamefont {Zhi}\ \bibnamefont {Ren}},
  \bibinfo {author} {\bibfnamefont {T.}~\bibnamefont {Sato}}, \bibinfo {author}
  {\bibfnamefont {K.}~\bibnamefont {Nakayama}}, \bibinfo {author}
  {\bibfnamefont {S.}~\bibnamefont {Souma}}, \bibinfo {author} {\bibfnamefont
  {T.}~\bibnamefont {Takahashi}}, \bibinfo {author} {\bibfnamefont {Kouji}\
  \bibnamefont {Segawa}}, \ and\ \bibinfo {author} {\bibfnamefont {Yoichi}\
  \bibnamefont {Ando}},\ }\bibfield  {title} {\enquote {\bibinfo {title}
  {Experimental realization of a topological crystalline insulator in snte},}\
  }\href {https://doi.org/10.1038/nphys2442} {\bibfield  {journal} {\bibinfo
  {journal} {Nat. Phys.}\ }\textbf {\bibinfo {volume} {8}},\ \bibinfo {pages}
  {800} (\bibinfo {year} {2012})}\BibitemShut {NoStop}%
\bibitem [{\citenamefont {Xu}\ \emph {et~al.}(2012)\citenamefont {Xu},
  \citenamefont {Liu}, \citenamefont {Alidoust}, \citenamefont {Neupane},
  \citenamefont {Qian}, \citenamefont {Belopolski}, \citenamefont {Denlinger},
  \citenamefont {Wang}, \citenamefont {Lin}, \citenamefont {Wray},
  \citenamefont {Landolt}, \citenamefont {Slomski}, \citenamefont {Dil},
  \citenamefont {Marcinkova}, \citenamefont {Morosan}, \citenamefont {Gibson},
  \citenamefont {Sankar}, \citenamefont {Chou}, \citenamefont {Cava},
  \citenamefont {Bansil},\ and\ \citenamefont {Hasan}}]{Xu2012}%
  \BibitemOpen
  \bibfield  {author} {\bibinfo {author} {\bibfnamefont {Su-Yang}\ \bibnamefont
  {Xu}}, \bibinfo {author} {\bibfnamefont {Chang}\ \bibnamefont {Liu}},
  \bibinfo {author} {\bibfnamefont {N.}~\bibnamefont {Alidoust}}, \bibinfo
  {author} {\bibfnamefont {M.}~\bibnamefont {Neupane}}, \bibinfo {author}
  {\bibfnamefont {D.}~\bibnamefont {Qian}}, \bibinfo {author} {\bibfnamefont
  {I.}~\bibnamefont {Belopolski}}, \bibinfo {author} {\bibfnamefont {J.~D.}\
  \bibnamefont {Denlinger}}, \bibinfo {author} {\bibfnamefont {Y.~J.}\
  \bibnamefont {Wang}}, \bibinfo {author} {\bibfnamefont {H.}~\bibnamefont
  {Lin}}, \bibinfo {author} {\bibfnamefont {L.~A.}\ \bibnamefont {Wray}},
  \bibinfo {author} {\bibfnamefont {G.}~\bibnamefont {Landolt}}, \bibinfo
  {author} {\bibfnamefont {B.}~\bibnamefont {Slomski}}, \bibinfo {author}
  {\bibfnamefont {J.~H.}\ \bibnamefont {Dil}}, \bibinfo {author} {\bibfnamefont
  {A.}~\bibnamefont {Marcinkova}}, \bibinfo {author} {\bibfnamefont
  {E.}~\bibnamefont {Morosan}}, \bibinfo {author} {\bibfnamefont
  {Q.}~\bibnamefont {Gibson}}, \bibinfo {author} {\bibfnamefont
  {R.}~\bibnamefont {Sankar}}, \bibinfo {author} {\bibfnamefont {F.~C.}\
  \bibnamefont {Chou}}, \bibinfo {author} {\bibfnamefont {R.~J.}\ \bibnamefont
  {Cava}}, \bibinfo {author} {\bibfnamefont {A.}~\bibnamefont {Bansil}}, \ and\
  \bibinfo {author} {\bibfnamefont {M.~Z.}\ \bibnamefont {Hasan}},\ }\bibfield
  {title} {\enquote {\bibinfo {title} {Observation of a topological crystalline
  insulator phase and topological phase transition in pb1-xsnxte},}\ }\href
  {https://doi.org/10.1038/ncomms2191} {\bibfield  {journal} {\bibinfo
  {journal} {Nat. Commun.}\ }\textbf {\bibinfo {volume} {3}},\ \bibinfo {pages}
  {1192} (\bibinfo {year} {2012})}\BibitemShut {NoStop}%
\bibitem [{\citenamefont {Neupane}\ \emph {et~al.}(2014)\citenamefont
  {Neupane}, \citenamefont {Xu}, \citenamefont {Sankar}, \citenamefont
  {Alidoust}, \citenamefont {Bian}, \citenamefont {Liu}, \citenamefont
  {Belopolski}, \citenamefont {Chang}, \citenamefont {Jeng}, \citenamefont
  {Lin}, \citenamefont {Bansil}, \citenamefont {Chou},\ and\ \citenamefont
  {Hasan}}]{Neupane2014}%
  \BibitemOpen
  \bibfield  {author} {\bibinfo {author} {\bibfnamefont {Madhab}\ \bibnamefont
  {Neupane}}, \bibinfo {author} {\bibfnamefont {Su-Yang}\ \bibnamefont {Xu}},
  \bibinfo {author} {\bibfnamefont {Raman}\ \bibnamefont {Sankar}}, \bibinfo
  {author} {\bibfnamefont {Nasser}\ \bibnamefont {Alidoust}}, \bibinfo {author}
  {\bibfnamefont {Guang}\ \bibnamefont {Bian}}, \bibinfo {author}
  {\bibfnamefont {Chang}\ \bibnamefont {Liu}}, \bibinfo {author} {\bibfnamefont
  {Ilya}\ \bibnamefont {Belopolski}}, \bibinfo {author} {\bibfnamefont
  {Tay-Rong}\ \bibnamefont {Chang}}, \bibinfo {author} {\bibfnamefont
  {Horng-Tay}\ \bibnamefont {Jeng}}, \bibinfo {author} {\bibfnamefont {Hsin}\
  \bibnamefont {Lin}}, \bibinfo {author} {\bibfnamefont {Arun}\ \bibnamefont
  {Bansil}}, \bibinfo {author} {\bibfnamefont {Fangcheng}\ \bibnamefont
  {Chou}}, \ and\ \bibinfo {author} {\bibfnamefont {M.~Zahid}\ \bibnamefont
  {Hasan}},\ }\bibfield  {title} {\enquote {\bibinfo {title} {Observation of a
  three-dimensional topological dirac semimetal phase in high-mobility
  cd3as2},}\ }\href {https://doi.org/10.1038/ncomms4786} {\bibfield  {journal}
  {\bibinfo  {journal} {Nat. Commun.}\ }\textbf {\bibinfo {volume} {5}},\
  \bibinfo {pages} {3786} (\bibinfo {year} {2014})}\BibitemShut {NoStop}%
\bibitem [{\citenamefont {Liu}\ \emph {et~al.}(2014)\citenamefont {Liu},
  \citenamefont {Zhou}, \citenamefont {Zhang}, \citenamefont {Wang},
  \citenamefont {Weng}, \citenamefont {Prabhakaran}, \citenamefont {Mo},
  \citenamefont {Shen}, \citenamefont {Fang}, \citenamefont {Dai},
  \citenamefont {Hussain},\ and\ \citenamefont {Chen}}]{Liu864}%
  \BibitemOpen
  \bibfield  {author} {\bibinfo {author} {\bibfnamefont {Z.~K.}\ \bibnamefont
  {Liu}}, \bibinfo {author} {\bibfnamefont {B.}~\bibnamefont {Zhou}}, \bibinfo
  {author} {\bibfnamefont {Y.}~\bibnamefont {Zhang}}, \bibinfo {author}
  {\bibfnamefont {Z.~J.}\ \bibnamefont {Wang}}, \bibinfo {author}
  {\bibfnamefont {H.~M.}\ \bibnamefont {Weng}}, \bibinfo {author}
  {\bibfnamefont {D.}~\bibnamefont {Prabhakaran}}, \bibinfo {author}
  {\bibfnamefont {S.-K.}\ \bibnamefont {Mo}}, \bibinfo {author} {\bibfnamefont
  {Z.~X.}\ \bibnamefont {Shen}}, \bibinfo {author} {\bibfnamefont
  {Z.}~\bibnamefont {Fang}}, \bibinfo {author} {\bibfnamefont {X.}~\bibnamefont
  {Dai}}, \bibinfo {author} {\bibfnamefont {Z.}~\bibnamefont {Hussain}}, \ and\
  \bibinfo {author} {\bibfnamefont {Y.~L.}\ \bibnamefont {Chen}},\ }\bibfield
  {title} {\enquote {\bibinfo {title} {Discovery of a three-dimensional
  topological dirac semimetal, na3bi},}\ }\href {\doibase
  10.1126/science.1245085} {\bibfield  {journal} {\bibinfo  {journal}
  {Science}\ }\textbf {\bibinfo {volume} {343}},\ \bibinfo {pages} {864--867}
  (\bibinfo {year} {2014})}\BibitemShut {NoStop}%
\bibitem [{\citenamefont {Lv}\ \emph {et~al.}(2015)\citenamefont {Lv},
  \citenamefont {Weng}, \citenamefont {Fu}, \citenamefont {Wang}, \citenamefont
  {Miao}, \citenamefont {Ma}, \citenamefont {Richard}, \citenamefont {Huang},
  \citenamefont {Zhao}, \citenamefont {Chen}, \citenamefont {Fang},
  \citenamefont {Dai}, \citenamefont {Qian},\ and\ \citenamefont
  {Ding}}]{TaAs_expt}%
  \BibitemOpen
  \bibfield  {author} {\bibinfo {author} {\bibfnamefont {B.~Q.}\ \bibnamefont
  {Lv}}, \bibinfo {author} {\bibfnamefont {H.~M.}\ \bibnamefont {Weng}},
  \bibinfo {author} {\bibfnamefont {B.~B.}\ \bibnamefont {Fu}}, \bibinfo
  {author} {\bibfnamefont {X.~P.}\ \bibnamefont {Wang}}, \bibinfo {author}
  {\bibfnamefont {H.}~\bibnamefont {Miao}}, \bibinfo {author} {\bibfnamefont
  {J.}~\bibnamefont {Ma}}, \bibinfo {author} {\bibfnamefont {P.}~\bibnamefont
  {Richard}}, \bibinfo {author} {\bibfnamefont {X.~C.}\ \bibnamefont {Huang}},
  \bibinfo {author} {\bibfnamefont {L.~X.}\ \bibnamefont {Zhao}}, \bibinfo
  {author} {\bibfnamefont {G.~F.}\ \bibnamefont {Chen}}, \bibinfo {author}
  {\bibfnamefont {Z.}~\bibnamefont {Fang}}, \bibinfo {author} {\bibfnamefont
  {X.}~\bibnamefont {Dai}}, \bibinfo {author} {\bibfnamefont {T.}~\bibnamefont
  {Qian}}, \ and\ \bibinfo {author} {\bibfnamefont {H.}~\bibnamefont {Ding}},\
  }\bibfield  {title} {\enquote {\bibinfo {title} {Experimental discovery of
  weyl semimetal taas},}\ }\href {\doibase 10.1103/PhysRevX.5.031013}
  {\bibfield  {journal} {\bibinfo  {journal} {Phys. Rev. X}\ }\textbf {\bibinfo
  {volume} {5}},\ \bibinfo {pages} {031013} (\bibinfo {year}
  {2015})}\BibitemShut {NoStop}%
\bibitem [{\citenamefont {Xu}\ \emph {et~al.}(2015)\citenamefont {Xu},
  \citenamefont {Belopolski}, \citenamefont {Alidoust}, \citenamefont
  {Neupane}, \citenamefont {Bian}, \citenamefont {Zhang}, \citenamefont
  {Sankar}, \citenamefont {Chang}, \citenamefont {Yuan}, \citenamefont {Lee},
  \citenamefont {Huang}, \citenamefont {Zheng}, \citenamefont {Ma},
  \citenamefont {Sanchez}, \citenamefont {Wang}, \citenamefont {Bansil},
  \citenamefont {Chou}, \citenamefont {Shibayev}, \citenamefont {Lin},
  \citenamefont {Jia},\ and\ \citenamefont {Hasan}}]{Xu613}%
  \BibitemOpen
  \bibfield  {author} {\bibinfo {author} {\bibfnamefont {Su-Yang}\ \bibnamefont
  {Xu}}, \bibinfo {author} {\bibfnamefont {Ilya}\ \bibnamefont {Belopolski}},
  \bibinfo {author} {\bibfnamefont {Nasser}\ \bibnamefont {Alidoust}}, \bibinfo
  {author} {\bibfnamefont {Madhab}\ \bibnamefont {Neupane}}, \bibinfo {author}
  {\bibfnamefont {Guang}\ \bibnamefont {Bian}}, \bibinfo {author}
  {\bibfnamefont {Chenglong}\ \bibnamefont {Zhang}}, \bibinfo {author}
  {\bibfnamefont {Raman}\ \bibnamefont {Sankar}}, \bibinfo {author}
  {\bibfnamefont {Guoqing}\ \bibnamefont {Chang}}, \bibinfo {author}
  {\bibfnamefont {Zhujun}\ \bibnamefont {Yuan}}, \bibinfo {author}
  {\bibfnamefont {Chi-Cheng}\ \bibnamefont {Lee}}, \bibinfo {author}
  {\bibfnamefont {Shin-Ming}\ \bibnamefont {Huang}}, \bibinfo {author}
  {\bibfnamefont {Hao}\ \bibnamefont {Zheng}}, \bibinfo {author} {\bibfnamefont
  {Jie}\ \bibnamefont {Ma}}, \bibinfo {author} {\bibfnamefont {Daniel~S.}\
  \bibnamefont {Sanchez}}, \bibinfo {author} {\bibfnamefont {BaoKai}\
  \bibnamefont {Wang}}, \bibinfo {author} {\bibfnamefont {Arun}\ \bibnamefont
  {Bansil}}, \bibinfo {author} {\bibfnamefont {Fangcheng}\ \bibnamefont
  {Chou}}, \bibinfo {author} {\bibfnamefont {Pavel~P.}\ \bibnamefont
  {Shibayev}}, \bibinfo {author} {\bibfnamefont {Hsin}\ \bibnamefont {Lin}},
  \bibinfo {author} {\bibfnamefont {Shuang}\ \bibnamefont {Jia}}, \ and\
  \bibinfo {author} {\bibfnamefont {M.~Zahid}\ \bibnamefont {Hasan}},\
  }\bibfield  {title} {\enquote {\bibinfo {title} {Discovery of a weyl fermion
  semimetal and topological fermi arcs},}\ }\href {\doibase
  10.1126/science.aaa9297} {\bibfield  {journal} {\bibinfo  {journal}
  {Science}\ }\textbf {\bibinfo {volume} {349}},\ \bibinfo {pages} {613--617}
  (\bibinfo {year} {2015})}\BibitemShut {NoStop}%
\bibitem [{\citenamefont {Bian}\ \emph {et~al.}(2016)\citenamefont {Bian},
  \citenamefont {Chang}, \citenamefont {Sankar}, \citenamefont {Xu},
  \citenamefont {Zheng}, \citenamefont {Neupert}, \citenamefont {Chiu},
  \citenamefont {Huang}, \citenamefont {Chang}, \citenamefont {Belopolski},
  \citenamefont {Sanchez}, \citenamefont {Neupane}, \citenamefont {Alidoust},
  \citenamefont {Liu}, \citenamefont {Wang}, \citenamefont {Lee}, \citenamefont
  {Jeng}, \citenamefont {Zhang}, \citenamefont {Yuan}, \citenamefont {Jia},
  \citenamefont {Bansil}, \citenamefont {Chou}, \citenamefont {Lin},\ and\
  \citenamefont {Hasan}}]{Bian2016}%
  \BibitemOpen
  \bibfield  {author} {\bibinfo {author} {\bibfnamefont {Guang}\ \bibnamefont
  {Bian}}, \bibinfo {author} {\bibfnamefont {Tay-Rong}\ \bibnamefont {Chang}},
  \bibinfo {author} {\bibfnamefont {Raman}\ \bibnamefont {Sankar}}, \bibinfo
  {author} {\bibfnamefont {Su-Yang}\ \bibnamefont {Xu}}, \bibinfo {author}
  {\bibfnamefont {Hao}\ \bibnamefont {Zheng}}, \bibinfo {author} {\bibfnamefont
  {Titus}\ \bibnamefont {Neupert}}, \bibinfo {author} {\bibfnamefont
  {Ching-Kai}\ \bibnamefont {Chiu}}, \bibinfo {author} {\bibfnamefont
  {Shin-Ming}\ \bibnamefont {Huang}}, \bibinfo {author} {\bibfnamefont
  {Guoqing}\ \bibnamefont {Chang}}, \bibinfo {author} {\bibfnamefont {Ilya}\
  \bibnamefont {Belopolski}}, \bibinfo {author} {\bibfnamefont {Daniel~S.}\
  \bibnamefont {Sanchez}}, \bibinfo {author} {\bibfnamefont {Madhab}\
  \bibnamefont {Neupane}}, \bibinfo {author} {\bibfnamefont {Nasser}\
  \bibnamefont {Alidoust}}, \bibinfo {author} {\bibfnamefont {Chang}\
  \bibnamefont {Liu}}, \bibinfo {author} {\bibfnamefont {BaoKai}\ \bibnamefont
  {Wang}}, \bibinfo {author} {\bibfnamefont {Chi-Cheng}\ \bibnamefont {Lee}},
  \bibinfo {author} {\bibfnamefont {Horng-Tay}\ \bibnamefont {Jeng}}, \bibinfo
  {author} {\bibfnamefont {Chenglong}\ \bibnamefont {Zhang}}, \bibinfo {author}
  {\bibfnamefont {Zhujun}\ \bibnamefont {Yuan}}, \bibinfo {author}
  {\bibfnamefont {Shuang}\ \bibnamefont {Jia}}, \bibinfo {author}
  {\bibfnamefont {Arun}\ \bibnamefont {Bansil}}, \bibinfo {author}
  {\bibfnamefont {Fangcheng}\ \bibnamefont {Chou}}, \bibinfo {author}
  {\bibfnamefont {Hsin}\ \bibnamefont {Lin}}, \ and\ \bibinfo {author}
  {\bibfnamefont {M.~Zahid}\ \bibnamefont {Hasan}},\ }\bibfield  {title}
  {\enquote {\bibinfo {title} {Topological nodal-line fermions in spin-orbit
  metal pbtase2},}\ }\href {http://dx.doi.org/10.1038/ncomms10556} {\bibfield
  {journal} {\bibinfo  {journal} {Nat. Commun.}\ }\textbf {\bibinfo {volume}
  {7}},\ \bibinfo {pages} {10556} (\bibinfo {year} {2016})}\BibitemShut
  {NoStop}%
\bibitem [{\citenamefont {Ma}\ \emph {et~al.}(2017)\citenamefont {Ma},
  \citenamefont {Yi}, \citenamefont {Lv}, \citenamefont {Wang}, \citenamefont
  {Nie}, \citenamefont {Wang}, \citenamefont {Kong}, \citenamefont {Huang},
  \citenamefont {Richard}, \citenamefont {Zhang}, \citenamefont {Yaji},
  \citenamefont {Kuroda}, \citenamefont {Shin}, \citenamefont {Weng},
  \citenamefont {Bernevig}, \citenamefont {Shi}, \citenamefont {Qian},\ and\
  \citenamefont {Ding}}]{Hourglass_expt}%
  \BibitemOpen
  \bibfield  {author} {\bibinfo {author} {\bibfnamefont {Junzhang}\
  \bibnamefont {Ma}}, \bibinfo {author} {\bibfnamefont {Changjiang}\
  \bibnamefont {Yi}}, \bibinfo {author} {\bibfnamefont {Baiqing}\ \bibnamefont
  {Lv}}, \bibinfo {author} {\bibfnamefont {ZhiJun}\ \bibnamefont {Wang}},
  \bibinfo {author} {\bibfnamefont {Simin}\ \bibnamefont {Nie}}, \bibinfo
  {author} {\bibfnamefont {Le}~\bibnamefont {Wang}}, \bibinfo {author}
  {\bibfnamefont {Lingyuan}\ \bibnamefont {Kong}}, \bibinfo {author}
  {\bibfnamefont {Yaobo}\ \bibnamefont {Huang}}, \bibinfo {author}
  {\bibfnamefont {Pierre}\ \bibnamefont {Richard}}, \bibinfo {author}
  {\bibfnamefont {Peng}\ \bibnamefont {Zhang}}, \bibinfo {author}
  {\bibfnamefont {Koichiro}\ \bibnamefont {Yaji}}, \bibinfo {author}
  {\bibfnamefont {Kenta}\ \bibnamefont {Kuroda}}, \bibinfo {author}
  {\bibfnamefont {Shik}\ \bibnamefont {Shin}}, \bibinfo {author} {\bibfnamefont
  {Hongming}\ \bibnamefont {Weng}}, \bibinfo {author} {\bibfnamefont
  {Bogdan~Andrei}\ \bibnamefont {Bernevig}}, \bibinfo {author} {\bibfnamefont
  {Youguo}\ \bibnamefont {Shi}}, \bibinfo {author} {\bibfnamefont {Tian}\
  \bibnamefont {Qian}}, \ and\ \bibinfo {author} {\bibfnamefont {Hong}\
  \bibnamefont {Ding}},\ }\bibfield  {title} {\enquote {\bibinfo {title}
  {Experimental evidence of hourglass fermion in the candidate nonsymmorphic
  topological insulator khgsb},}\ }\href {\doibase
  https://doi.org/10.1126/sciadv.1602415} {\bibfield  {journal} {\bibinfo
  {journal} {Sci. Adv.}\ }\textbf {\bibinfo {volume} {3}},\ \bibinfo {pages}
  {e1602415} (\bibinfo {year} {2017})}\BibitemShut {NoStop}%
\bibitem [{\citenamefont {Rauch}\ \emph {et~al.}(2014)\citenamefont {Rauch},
  \citenamefont {Flieger}, \citenamefont {Henk}, \citenamefont {Mertig},\ and\
  \citenamefont {Ernst}}]{Bi2Te3_dualtopology}%
  \BibitemOpen
  \bibfield  {author} {\bibinfo {author} {\bibfnamefont {Tom\'a\ifmmode
  \check{s}\else~\v{s}\fi{}}\ \bibnamefont {Rauch}}, \bibinfo {author}
  {\bibfnamefont {Markus}\ \bibnamefont {Flieger}}, \bibinfo {author}
  {\bibfnamefont {J\"urgen}\ \bibnamefont {Henk}}, \bibinfo {author}
  {\bibfnamefont {Ingrid}\ \bibnamefont {Mertig}}, \ and\ \bibinfo {author}
  {\bibfnamefont {Arthur}\ \bibnamefont {Ernst}},\ }\bibfield  {title}
  {\enquote {\bibinfo {title} {Dual topological character of chalcogenides:
  Theory for ${\mathrm{bi}}_{2}{\mathrm{te}}_{3}$},}\ }\href {\doibase
  10.1103/PhysRevLett.112.016802} {\bibfield  {journal} {\bibinfo  {journal}
  {Phys. Rev. Lett.}\ }\textbf {\bibinfo {volume} {112}},\ \bibinfo {pages}
  {016802} (\bibinfo {year} {2014})}\BibitemShut {NoStop}%
\bibitem [{\citenamefont {Eschbach}\ \emph {et~al.}(2017)\citenamefont
  {Eschbach}, \citenamefont {Lanius}, \citenamefont {Niu}, \citenamefont
  {Mlynczak}, \citenamefont {Gospodaric}, \citenamefont {Kellner},
  \citenamefont {Sch{\"u}ffelgen}, \citenamefont {Gehlmann}, \citenamefont
  {D{\"o}ring}, \citenamefont {Neumann}, \citenamefont {Luysberg},
  \citenamefont {Mussler}, \citenamefont {Plucinski}, \citenamefont
  {Morgenstern}, \citenamefont {Gr{\"u}tzmacher}, \citenamefont {Bihlmayer},
  \citenamefont {Bl{\"u}gel},\ and\ \citenamefont {Schneider}}]{Bi1Te1_dual}%
  \BibitemOpen
  \bibfield  {author} {\bibinfo {author} {\bibfnamefont {Markus}\ \bibnamefont
  {Eschbach}}, \bibinfo {author} {\bibfnamefont {Martin}\ \bibnamefont
  {Lanius}}, \bibinfo {author} {\bibfnamefont {Chengwang}\ \bibnamefont {Niu}},
  \bibinfo {author} {\bibfnamefont {Ewa}\ \bibnamefont {Mlynczak}}, \bibinfo
  {author} {\bibfnamefont {Pika}\ \bibnamefont {Gospodaric}}, \bibinfo {author}
  {\bibfnamefont {Jens}\ \bibnamefont {Kellner}}, \bibinfo {author}
  {\bibfnamefont {Peter}\ \bibnamefont {Sch{\"u}ffelgen}}, \bibinfo {author}
  {\bibfnamefont {Mathias}\ \bibnamefont {Gehlmann}}, \bibinfo {author}
  {\bibfnamefont {Sven}\ \bibnamefont {D{\"o}ring}}, \bibinfo {author}
  {\bibfnamefont {Elmar}\ \bibnamefont {Neumann}}, \bibinfo {author}
  {\bibfnamefont {Martina}\ \bibnamefont {Luysberg}}, \bibinfo {author}
  {\bibfnamefont {Gregor}\ \bibnamefont {Mussler}}, \bibinfo {author}
  {\bibfnamefont {Lukasz}\ \bibnamefont {Plucinski}}, \bibinfo {author}
  {\bibfnamefont {Markus}\ \bibnamefont {Morgenstern}}, \bibinfo {author}
  {\bibfnamefont {Detlev}\ \bibnamefont {Gr{\"u}tzmacher}}, \bibinfo {author}
  {\bibfnamefont {Gustav}\ \bibnamefont {Bihlmayer}}, \bibinfo {author}
  {\bibfnamefont {Stefan}\ \bibnamefont {Bl{\"u}gel}}, \ and\ \bibinfo {author}
  {\bibfnamefont {Claus~M.}\ \bibnamefont {Schneider}},\ }\bibfield  {title}
  {\enquote {\bibinfo {title} {Bi1te1 is a dual topological insulator},}\
  }\href {https://doi.org/10.1038/ncomms14976} {\bibfield  {journal} {\bibinfo
  {journal} {Nat. Commun.}\ }\textbf {\bibinfo {volume} {8}},\ \bibinfo {pages}
  {14976} (\bibinfo {year} {2017})}\BibitemShut {NoStop}%
\bibitem [{\citenamefont {Singh}\ \emph
  {et~al.}(2018{\natexlab{b}})\citenamefont {Singh}, \citenamefont {Zhou},
  \citenamefont {Lin},\ and\ \citenamefont {Bansil}}]{singh18_saddle}%
  \BibitemOpen
  \bibfield  {author} {\bibinfo {author} {\bibfnamefont {Bahadur}\ \bibnamefont
  {Singh}}, \bibinfo {author} {\bibfnamefont {Xiaoting}\ \bibnamefont {Zhou}},
  \bibinfo {author} {\bibfnamefont {Hsin}\ \bibnamefont {Lin}}, \ and\ \bibinfo
  {author} {\bibfnamefont {Arun}\ \bibnamefont {Bansil}},\ }\bibfield  {title}
  {\enquote {\bibinfo {title} {Saddle-like topological surface states on the
  $t{T}^{\ensuremath{'}}x$ family of compounds ($t, {T}^{\ensuremath{'}}$ =
  transition metal, $x=\mathrm{Si}$, ge)},}\ }\href {\doibase
  10.1103/PhysRevB.97.075125} {\bibfield  {journal} {\bibinfo  {journal} {Phys.
  Rev. B}\ }\textbf {\bibinfo {volume} {97}},\ \bibinfo {pages} {075125}
  (\bibinfo {year} {2018}{\natexlab{b}})}\BibitemShut {NoStop}%
\bibitem [{\citenamefont {Cao}\ \emph {et~al.}(2018{\natexlab{a}})\citenamefont
  {Cao}, \citenamefont {Fatemi}, \citenamefont {Demir}, \citenamefont {Fang},
  \citenamefont {Tomarken}, \citenamefont {Luo}, \citenamefont
  {Sanchez-Yamagishi}, \citenamefont {Watanabe}, \citenamefont {Taniguchi},
  \citenamefont {Kaxiras}, \citenamefont {Ashoori},\ and\ \citenamefont
  {Jarillo-Herrero}}]{TBLG_correlated_insulator}%
  \BibitemOpen
  \bibfield  {author} {\bibinfo {author} {\bibfnamefont {Yuan}\ \bibnamefont
  {Cao}}, \bibinfo {author} {\bibfnamefont {Valla}\ \bibnamefont {Fatemi}},
  \bibinfo {author} {\bibfnamefont {Ahmet}\ \bibnamefont {Demir}}, \bibinfo
  {author} {\bibfnamefont {Shiang}\ \bibnamefont {Fang}}, \bibinfo {author}
  {\bibfnamefont {Spencer~L.}\ \bibnamefont {Tomarken}}, \bibinfo {author}
  {\bibfnamefont {Jason~Y.}\ \bibnamefont {Luo}}, \bibinfo {author}
  {\bibfnamefont {Javier~D.}\ \bibnamefont {Sanchez-Yamagishi}}, \bibinfo
  {author} {\bibfnamefont {Kenji}\ \bibnamefont {Watanabe}}, \bibinfo {author}
  {\bibfnamefont {Takashi}\ \bibnamefont {Taniguchi}}, \bibinfo {author}
  {\bibfnamefont {Efthimios}\ \bibnamefont {Kaxiras}}, \bibinfo {author}
  {\bibfnamefont {Ray~C.}\ \bibnamefont {Ashoori}}, \ and\ \bibinfo {author}
  {\bibfnamefont {Pablo}\ \bibnamefont {Jarillo-Herrero}},\ }\bibfield  {title}
  {\enquote {\bibinfo {title} {Correlated insulator behaviour at half-filling
  in magic-angle graphene superlattices},}\ }\href
  {https://doi.org/10.1038/nature26154} {\bibfield  {journal} {\bibinfo
  {journal} {Nature}\ }\textbf {\bibinfo {volume} {556}},\ \bibinfo {pages}
  {80} (\bibinfo {year} {2018}{\natexlab{a}})}\BibitemShut {NoStop}%
\bibitem [{\citenamefont {Cao}\ \emph {et~al.}(2018{\natexlab{b}})\citenamefont
  {Cao}, \citenamefont {Fatemi}, \citenamefont {Fang}, \citenamefont
  {Watanabe}, \citenamefont {Taniguchi}, \citenamefont {Kaxiras},\ and\
  \citenamefont {Jarillo-Herrero}}]{TBLG_superconductivity}%
  \BibitemOpen
  \bibfield  {author} {\bibinfo {author} {\bibfnamefont {Yuan}\ \bibnamefont
  {Cao}}, \bibinfo {author} {\bibfnamefont {Valla}\ \bibnamefont {Fatemi}},
  \bibinfo {author} {\bibfnamefont {Shiang}\ \bibnamefont {Fang}}, \bibinfo
  {author} {\bibfnamefont {Kenji}\ \bibnamefont {Watanabe}}, \bibinfo {author}
  {\bibfnamefont {Takashi}\ \bibnamefont {Taniguchi}}, \bibinfo {author}
  {\bibfnamefont {Efthimios}\ \bibnamefont {Kaxiras}}, \ and\ \bibinfo {author}
  {\bibfnamefont {Pablo}\ \bibnamefont {Jarillo-Herrero}},\ }\bibfield  {title}
  {\enquote {\bibinfo {title} {Unconventional superconductivity in magic-angle
  graphene superlattices},}\ }\href {https://doi.org/10.1038/nature26160}
  {\bibfield  {journal} {\bibinfo  {journal} {Nature}\ }\textbf {\bibinfo
  {volume} {556}},\ \bibinfo {pages} {43} (\bibinfo {year}
  {2018}{\natexlab{b}})},\ \bibinfo {note} {article}\BibitemShut {NoStop}%
\bibitem [{\citenamefont {Yuan}\ \emph {et~al.}(2019)\citenamefont {Yuan},
  \citenamefont {Isobe},\ and\ \citenamefont {Fu}}]{LiangFu_higher_order_VHS}%
  \BibitemOpen
  \bibfield  {author} {\bibinfo {author} {\bibfnamefont {Noah F.~Q.}\
  \bibnamefont {Yuan}}, \bibinfo {author} {\bibfnamefont {Hiroki}\ \bibnamefont
  {Isobe}}, \ and\ \bibinfo {author} {\bibfnamefont {Liang}\ \bibnamefont
  {Fu}},\ }\bibfield  {title} {\enquote {\bibinfo {title} {Magic of high-order
  van hove singularity},}\ }\href {\doibase 10.1038/s41467-019-13670-9}
  {\bibfield  {journal} {\bibinfo  {journal} {Nature Communications}\ }\textbf
  {\bibinfo {volume} {10}},\ \bibinfo {pages} {5769} (\bibinfo {year}
  {2019})}\BibitemShut {NoStop}%
\bibitem [{\citenamefont {Hlubina}\ \emph {et~al.}(1997)\citenamefont
  {Hlubina}, \citenamefont {Sorella},\ and\ \citenamefont
  {Guinea}}]{PhysRevLett.78.1343}%
  \BibitemOpen
  \bibfield  {author} {\bibinfo {author} {\bibfnamefont {R.}~\bibnamefont
  {Hlubina}}, \bibinfo {author} {\bibfnamefont {S.}~\bibnamefont {Sorella}}, \
  and\ \bibinfo {author} {\bibfnamefont {F.}~\bibnamefont {Guinea}},\
  }\bibfield  {title} {\enquote {\bibinfo {title} {Ferromagnetism in the two
  dimensional $t\ensuremath{-}{t}^{\ensuremath{'}}$ hubbard model at the van
  hove density},}\ }\href {\doibase 10.1103/PhysRevLett.78.1343} {\bibfield
  {journal} {\bibinfo  {journal} {Phys. Rev. Lett.}\ }\textbf {\bibinfo
  {volume} {78}},\ \bibinfo {pages} {1343--1346} (\bibinfo {year}
  {1997})}\BibitemShut {NoStop}%
\bibitem [{\citenamefont {Honerkamp}\ and\ \citenamefont
  {Salmhofer}(2001)}]{PhysRevLett.87.187004}%
  \BibitemOpen
  \bibfield  {author} {\bibinfo {author} {\bibfnamefont {Carsten}\ \bibnamefont
  {Honerkamp}}\ and\ \bibinfo {author} {\bibfnamefont {Manfred}\ \bibnamefont
  {Salmhofer}},\ }\bibfield  {title} {\enquote {\bibinfo {title} {Magnetic and
  superconducting instabilities of the hubbard model at the van hove
  filling},}\ }\href {\doibase 10.1103/PhysRevLett.87.187004} {\bibfield
  {journal} {\bibinfo  {journal} {Phys. Rev. Lett.}\ }\textbf {\bibinfo
  {volume} {87}},\ \bibinfo {pages} {187004} (\bibinfo {year}
  {2001})}\BibitemShut {NoStop}%
\bibitem [{\citenamefont {Ziletti}\ \emph {et~al.}(2015)\citenamefont
  {Ziletti}, \citenamefont {Huang}, \citenamefont {Coker},\ and\ \citenamefont
  {Lin}}]{PhysRevB.92.085423}%
  \BibitemOpen
  \bibfield  {author} {\bibinfo {author} {\bibfnamefont {A.}~\bibnamefont
  {Ziletti}}, \bibinfo {author} {\bibfnamefont {S.~M.}\ \bibnamefont {Huang}},
  \bibinfo {author} {\bibfnamefont {D.~F.}\ \bibnamefont {Coker}}, \ and\
  \bibinfo {author} {\bibfnamefont {H.}~\bibnamefont {Lin}},\ }\bibfield
  {title} {\enquote {\bibinfo {title} {Van hove singularity and ferromagnetic
  instability in phosphorene},}\ }\href {\doibase 10.1103/PhysRevB.92.085423}
  {\bibfield  {journal} {\bibinfo  {journal} {Phys. Rev. B}\ }\textbf {\bibinfo
  {volume} {92}},\ \bibinfo {pages} {085423} (\bibinfo {year}
  {2015})}\BibitemShut {NoStop}%
\bibitem [{\citenamefont {Singh}\ \emph {et~al.}(2017)\citenamefont {Singh},
  \citenamefont {Hsu}, \citenamefont {Tsai}, \citenamefont {Pereira},\ and\
  \citenamefont {Lin}}]{PhysRevB.95.245136}%
  \BibitemOpen
  \bibfield  {author} {\bibinfo {author} {\bibfnamefont {Bahadur}\ \bibnamefont
  {Singh}}, \bibinfo {author} {\bibfnamefont {Chuang-Han}\ \bibnamefont {Hsu}},
  \bibinfo {author} {\bibfnamefont {Wei-Feng}\ \bibnamefont {Tsai}}, \bibinfo
  {author} {\bibfnamefont {Vitor~M.}\ \bibnamefont {Pereira}}, \ and\ \bibinfo
  {author} {\bibfnamefont {Hsin}\ \bibnamefont {Lin}},\ }\bibfield  {title}
  {\enquote {\bibinfo {title} {Stable charge density wave phase in a
  $1t--{\mathrm{tise}}_{2}$ monolayer},}\ }\href {\doibase
  10.1103/PhysRevB.95.245136} {\bibfield  {journal} {\bibinfo  {journal} {Phys.
  Rev. B}\ }\textbf {\bibinfo {volume} {95}},\ \bibinfo {pages} {245136}
  (\bibinfo {year} {2017})}\BibitemShut {NoStop}%
\bibitem [{\citenamefont {Chen}\ and\ \citenamefont
  {Lado}(2019)}]{DSS_surface_FM}%
  \BibitemOpen
  \bibfield  {author} {\bibinfo {author} {\bibfnamefont {Wei}\ \bibnamefont
  {Chen}}\ and\ \bibinfo {author} {\bibfnamefont {J.~L.}\ \bibnamefont
  {Lado}},\ }\bibfield  {title} {\enquote {\bibinfo {title} {Interaction-driven
  surface chern insulator in nodal line semimetals},}\ }\href {\doibase
  10.1103/PhysRevLett.122.016803} {\bibfield  {journal} {\bibinfo  {journal}
  {Phys. Rev. Lett.}\ }\textbf {\bibinfo {volume} {122}},\ \bibinfo {pages}
  {016803} (\bibinfo {year} {2019})}\BibitemShut {NoStop}%
\bibitem [{\citenamefont {Yao}\ and\ \citenamefont
  {Yang}(2015)}]{VHS_superconductivity}%
  \BibitemOpen
  \bibfield  {author} {\bibinfo {author} {\bibfnamefont {Hong}\ \bibnamefont
  {Yao}}\ and\ \bibinfo {author} {\bibfnamefont {Fan}\ \bibnamefont {Yang}},\
  }\bibfield  {title} {\enquote {\bibinfo {title} {Topological odd-parity
  superconductivity at type-ii two-dimensional van hove singularities},}\
  }\href {\doibase 10.1103/PhysRevB.92.035132} {\bibfield  {journal} {\bibinfo
  {journal} {Phys. Rev. B}\ }\textbf {\bibinfo {volume} {92}},\ \bibinfo
  {pages} {035132} (\bibinfo {year} {2015})}\BibitemShut {NoStop}%
\bibitem [{\citenamefont {Markiewicz}(1997)}]{MARKIEWICZ_VHS_SC}%
  \BibitemOpen
  \bibfield  {author} {\bibinfo {author} {\bibfnamefont {R.S}\ \bibnamefont
  {Markiewicz}},\ }\bibfield  {title} {\enquote {\bibinfo {title} {A survey of
  the van hove scenario for high-tc superconductivity with special emphasis on
  pseudogaps and striped phases},}\ }\href {\doibase
  https://doi.org/10.1016/S0022-3697(97)00025-5} {\bibfield  {journal}
  {\bibinfo  {journal} {J. Phys. Chem. Solids}\ }\textbf {\bibinfo {volume}
  {58}},\ \bibinfo {pages} {1179 -- 1310} (\bibinfo {year} {1997})}\BibitemShut
  {NoStop}%
\bibitem [{\citenamefont {Marrazzo}\ \emph {et~al.}(2018)\citenamefont
  {Marrazzo}, \citenamefont {Gibertini}, \citenamefont {Campi}, \citenamefont
  {Mounet},\ and\ \citenamefont {Marzari}}]{HgPt2Se3_monolayer}%
  \BibitemOpen
  \bibfield  {author} {\bibinfo {author} {\bibfnamefont {Antimo}\ \bibnamefont
  {Marrazzo}}, \bibinfo {author} {\bibfnamefont {Marco}\ \bibnamefont
  {Gibertini}}, \bibinfo {author} {\bibfnamefont {Davide}\ \bibnamefont
  {Campi}}, \bibinfo {author} {\bibfnamefont {Nicolas}\ \bibnamefont {Mounet}},
  \ and\ \bibinfo {author} {\bibfnamefont {Nicola}\ \bibnamefont {Marzari}},\
  }\bibfield  {title} {\enquote {\bibinfo {title} {Prediction of a large-gap
  and switchable kane-mele quantum spin hall insulator},}\ }\href {\doibase
  10.1103/PhysRevLett.120.117701} {\bibfield  {journal} {\bibinfo  {journal}
  {Phys. Rev. Lett.}\ }\textbf {\bibinfo {volume} {120}},\ \bibinfo {pages}
  {117701} (\bibinfo {year} {2018})}\BibitemShut {NoStop}%
\bibitem [{\citenamefont {{Wu}}\ \emph {et~al.}(2018)\citenamefont {{Wu}},
  \citenamefont {{Fink}}, \citenamefont {{Hanke}}, \citenamefont {{Thomale}},\
  and\ \citenamefont {{Di Sante}}}]{HgPt2Se3_superconductivity}%
  \BibitemOpen
  \bibfield  {author} {\bibinfo {author} {\bibfnamefont {Xianxin}\ \bibnamefont
  {{Wu}}}, \bibinfo {author} {\bibfnamefont {Mario}\ \bibnamefont {{Fink}}},
  \bibinfo {author} {\bibfnamefont {Werner}\ \bibnamefont {{Hanke}}}, \bibinfo
  {author} {\bibfnamefont {Ronny}\ \bibnamefont {{Thomale}}}, \ and\ \bibinfo
  {author} {\bibfnamefont {Domenico}\ \bibnamefont {{Di Sante}}},\ }\bibfield
  {title} {\enquote {\bibinfo {title} {{Unconventional superconductivity in a
  doped quantum spin Hall insulator}},}\ }\href@noop {} {\ ,\ \bibinfo {eid}
  {arXiv:1811.01746} (\bibinfo {year} {2018})},\ \Eprint
  {http://arxiv.org/abs/1811.01746} {arXiv:1811.01746 [cond-mat.supr-con]}
  \BibitemShut {NoStop}%
\bibitem [{\citenamefont {{Kandrai}}\ \emph {et~al.}(2019)\citenamefont
  {{Kandrai}}, \citenamefont {{Kukucska}}, \citenamefont {{Vancs{\'o}}},
  \citenamefont {{Koltai}}, \citenamefont {{Baranka}}, \citenamefont
  {{Horv{\'a}th}}, \citenamefont {{Hoffmann}}, \citenamefont
  {{Vymazalov{\'a}}}, \citenamefont {{Tapaszt{\'o}}},\ and\ \citenamefont
  {{Nemes-Incze}}}]{HgPt2Se3_expt}%
  \BibitemOpen
  \bibfield  {author} {\bibinfo {author} {\bibfnamefont {Konr{\'a}d}\
  \bibnamefont {{Kandrai}}}, \bibinfo {author} {\bibfnamefont {Gerg{\H{o}}}\
  \bibnamefont {{Kukucska}}}, \bibinfo {author} {\bibfnamefont {P{\'e}ter}\
  \bibnamefont {{Vancs{\'o}}}}, \bibinfo {author} {\bibfnamefont {J{\'a}nos}\
  \bibnamefont {{Koltai}}}, \bibinfo {author} {\bibfnamefont {Gy{\"o}rgy}\
  \bibnamefont {{Baranka}}}, \bibinfo {author} {\bibfnamefont {Zsolt~E.}\
  \bibnamefont {{Horv{\'a}th}}}, \bibinfo {author} {\bibfnamefont {{\'A}kos}\
  \bibnamefont {{Hoffmann}}}, \bibinfo {author} {\bibfnamefont {Anna}\
  \bibnamefont {{Vymazalov{\'a}}}}, \bibinfo {author} {\bibfnamefont {Levente}\
  \bibnamefont {{Tapaszt{\'o}}}}, \ and\ \bibinfo {author} {\bibfnamefont
  {P{\'e}ter}\ \bibnamefont {{Nemes-Incze}}},\ }\bibfield  {title} {\enquote
  {\bibinfo {title} {{Evidence for room temperature quantum spin Hall state in
  the layered mineral jacutingaite}},}\ }\href@noop {} {\ ,\ \bibinfo {eid}
  {arXiv:1903.02458} (\bibinfo {year} {2019})},\ \Eprint
  {http://arxiv.org/abs/1903.02458} {arXiv:1903.02458 [cond-mat.mes-hall]}
  \BibitemShut {NoStop}%
\bibitem [{\citenamefont {Kohn}\ and\ \citenamefont
  {Sham}(1965)}]{PhysRev.140.A1133}%
  \BibitemOpen
  \bibfield  {author} {\bibinfo {author} {\bibfnamefont {W.}~\bibnamefont
  {Kohn}}\ and\ \bibinfo {author} {\bibfnamefont {L.~J.}\ \bibnamefont
  {Sham}},\ }\bibfield  {title} {\enquote {\bibinfo {title} {Self-consistent
  equations including exchange and correlation effects},}\ }\href {\doibase
  10.1103/PhysRev.140.A1133} {\bibfield  {journal} {\bibinfo  {journal} {Phys.
  Rev.}\ }\textbf {\bibinfo {volume} {140}},\ \bibinfo {pages} {A1133--A1138}
  (\bibinfo {year} {1965})}\BibitemShut {NoStop}%
\bibitem [{\citenamefont {Kresse}\ and\ \citenamefont
  {Joubert}(1999)}]{PhysRevB.59.1758}%
  \BibitemOpen
  \bibfield  {author} {\bibinfo {author} {\bibfnamefont {G.}~\bibnamefont
  {Kresse}}\ and\ \bibinfo {author} {\bibfnamefont {D.}~\bibnamefont
  {Joubert}},\ }\bibfield  {title} {\enquote {\bibinfo {title} {From ultrasoft
  pseudopotentials to the projector augmented-wave method},}\ }\href {\doibase
  10.1103/PhysRevB.59.1758} {\bibfield  {journal} {\bibinfo  {journal} {Phys.
  Rev. B}\ }\textbf {\bibinfo {volume} {59}},\ \bibinfo {pages} {1758--1775}
  (\bibinfo {year} {1999})}\BibitemShut {NoStop}%
\bibitem [{\citenamefont {Giannozzi}\ \emph {et~al.}(2009)\citenamefont
  {Giannozzi}, \citenamefont {Baroni}, \citenamefont {Bonini}, \citenamefont
  {Calandra}, \citenamefont {Car}, \citenamefont {Cavazzoni}, \citenamefont
  {Ceresoli}, \citenamefont {Chiarotti}, \citenamefont {Cococcioni},
  \citenamefont {Dabo}, \citenamefont {Corso}, \citenamefont {de~Gironcoli},
  \citenamefont {Fabris}, \citenamefont {Fratesi}, \citenamefont {Gebauer},
  \citenamefont {Gerstmann}, \citenamefont {Gougoussis}, \citenamefont
  {Kokalj}, \citenamefont {Lazzeri}, \citenamefont {Martin-Samos},
  \citenamefont {Marzari}, \citenamefont {Mauri}, \citenamefont {Mazzarello},
  \citenamefont {Paolini}, \citenamefont {Pasquarello}, \citenamefont
  {Paulatto}, \citenamefont {Sbraccia}, \citenamefont {Scandolo}, \citenamefont
  {Sclauzero}, \citenamefont {Seitsonen}, \citenamefont {Smogunov},
  \citenamefont {Umari},\ and\ \citenamefont {Wentzcovitch}}]{Giannozzi_2009}%
  \BibitemOpen
  \bibfield  {author} {\bibinfo {author} {\bibfnamefont {Paolo}\ \bibnamefont
  {Giannozzi}}, \bibinfo {author} {\bibfnamefont {Stefano}\ \bibnamefont
  {Baroni}}, \bibinfo {author} {\bibfnamefont {Nicola}\ \bibnamefont {Bonini}},
  \bibinfo {author} {\bibfnamefont {Matteo}\ \bibnamefont {Calandra}}, \bibinfo
  {author} {\bibfnamefont {Roberto}\ \bibnamefont {Car}}, \bibinfo {author}
  {\bibfnamefont {Carlo}\ \bibnamefont {Cavazzoni}}, \bibinfo {author}
  {\bibfnamefont {Davide}\ \bibnamefont {Ceresoli}}, \bibinfo {author}
  {\bibfnamefont {Guido~L}\ \bibnamefont {Chiarotti}}, \bibinfo {author}
  {\bibfnamefont {Matteo}\ \bibnamefont {Cococcioni}}, \bibinfo {author}
  {\bibfnamefont {Ismaila}\ \bibnamefont {Dabo}}, \bibinfo {author}
  {\bibfnamefont {Andrea~Dal}\ \bibnamefont {Corso}}, \bibinfo {author}
  {\bibfnamefont {Stefano}\ \bibnamefont {de~Gironcoli}}, \bibinfo {author}
  {\bibfnamefont {Stefano}\ \bibnamefont {Fabris}}, \bibinfo {author}
  {\bibfnamefont {Guido}\ \bibnamefont {Fratesi}}, \bibinfo {author}
  {\bibfnamefont {Ralph}\ \bibnamefont {Gebauer}}, \bibinfo {author}
  {\bibfnamefont {Uwe}\ \bibnamefont {Gerstmann}}, \bibinfo {author}
  {\bibfnamefont {Christos}\ \bibnamefont {Gougoussis}}, \bibinfo {author}
  {\bibfnamefont {Anton}\ \bibnamefont {Kokalj}}, \bibinfo {author}
  {\bibfnamefont {Michele}\ \bibnamefont {Lazzeri}}, \bibinfo {author}
  {\bibfnamefont {Layla}\ \bibnamefont {Martin-Samos}}, \bibinfo {author}
  {\bibfnamefont {Nicola}\ \bibnamefont {Marzari}}, \bibinfo {author}
  {\bibfnamefont {Francesco}\ \bibnamefont {Mauri}}, \bibinfo {author}
  {\bibfnamefont {Riccardo}\ \bibnamefont {Mazzarello}}, \bibinfo {author}
  {\bibfnamefont {Stefano}\ \bibnamefont {Paolini}}, \bibinfo {author}
  {\bibfnamefont {Alfredo}\ \bibnamefont {Pasquarello}}, \bibinfo {author}
  {\bibfnamefont {Lorenzo}\ \bibnamefont {Paulatto}}, \bibinfo {author}
  {\bibfnamefont {Carlo}\ \bibnamefont {Sbraccia}}, \bibinfo {author}
  {\bibfnamefont {Sandro}\ \bibnamefont {Scandolo}}, \bibinfo {author}
  {\bibfnamefont {Gabriele}\ \bibnamefont {Sclauzero}}, \bibinfo {author}
  {\bibfnamefont {Ari~P}\ \bibnamefont {Seitsonen}}, \bibinfo {author}
  {\bibfnamefont {Alexander}\ \bibnamefont {Smogunov}}, \bibinfo {author}
  {\bibfnamefont {Paolo}\ \bibnamefont {Umari}}, \ and\ \bibinfo {author}
  {\bibfnamefont {Renata~M}\ \bibnamefont {Wentzcovitch}},\ }\bibfield  {title}
  {\enquote {\bibinfo {title} {{QUANTUM} {ESPRESSO}: a modular and open-source
  software project for quantum simulations of materials},}\ }\href {\doibase
  10.1088/0953-8984/21/39/395502} {\bibfield  {journal} {\bibinfo  {journal}
  {J. Phys. Condens. Matter}\ }\textbf {\bibinfo {volume} {21}},\ \bibinfo
  {pages} {395502} (\bibinfo {year} {2009})}\BibitemShut {NoStop}%
\bibitem [{\citenamefont {Perdew}\ \emph {et~al.}(1996)\citenamefont {Perdew},
  \citenamefont {Burke},\ and\ \citenamefont
  {Ernzerhof}}]{PhysRevLett.77.3865}%
  \BibitemOpen
  \bibfield  {author} {\bibinfo {author} {\bibfnamefont {John~P.}\ \bibnamefont
  {Perdew}}, \bibinfo {author} {\bibfnamefont {Kieron}\ \bibnamefont {Burke}},
  \ and\ \bibinfo {author} {\bibfnamefont {Matthias}\ \bibnamefont
  {Ernzerhof}},\ }\bibfield  {title} {\enquote {\bibinfo {title} {Generalized
  gradient approximation made simple},}\ }\href {\doibase
  10.1103/PhysRevLett.77.3865} {\bibfield  {journal} {\bibinfo  {journal}
  {Phys. Rev. Lett.}\ }\textbf {\bibinfo {volume} {77}},\ \bibinfo {pages}
  {3865--3868} (\bibinfo {year} {1996})}\BibitemShut {NoStop}%
\bibitem [{\citenamefont {Grimme}\ \emph {et~al.}(2011)\citenamefont {Grimme},
  \citenamefont {Ehrlich},\ and\ \citenamefont {Goerigk}}]{GrimmieDFTD3BJ}%
  \BibitemOpen
  \bibfield  {author} {\bibinfo {author} {\bibfnamefont {Stefan}\ \bibnamefont
  {Grimme}}, \bibinfo {author} {\bibfnamefont {Stephan}\ \bibnamefont
  {Ehrlich}}, \ and\ \bibinfo {author} {\bibfnamefont {Lars}\ \bibnamefont
  {Goerigk}},\ }\bibfield  {title} {\enquote {\bibinfo {title} {Effect of the
  damping function in dispersion corrected density functional theory},}\ }\href
  {\doibase 10.1002/jcc.21759} {\bibfield  {journal} {\bibinfo  {journal}
  {Journal of Computational Chemistry}\ }\textbf {\bibinfo {volume} {32}},\
  \bibinfo {pages} {1456--1465} (\bibinfo {year} {2011})}\BibitemShut {NoStop}%
\bibitem [{\citenamefont {Marzari}\ and\ \citenamefont
  {Vanderbilt}(1997)}]{PhysRevB.56.12847}%
  \BibitemOpen
  \bibfield  {author} {\bibinfo {author} {\bibfnamefont {Nicola}\ \bibnamefont
  {Marzari}}\ and\ \bibinfo {author} {\bibfnamefont {David}\ \bibnamefont
  {Vanderbilt}},\ }\bibfield  {title} {\enquote {\bibinfo {title} {Maximally
  localized generalized wannier functions for composite energy bands},}\ }\href
  {\doibase 10.1103/PhysRevB.56.12847} {\bibfield  {journal} {\bibinfo
  {journal} {Phys. Rev. B}\ }\textbf {\bibinfo {volume} {56}},\ \bibinfo
  {pages} {12847--12865} (\bibinfo {year} {1997})}\BibitemShut {NoStop}%
\bibitem [{\citenamefont {Wu}\ \emph {et~al.}(2018)\citenamefont {Wu},
  \citenamefont {Zhang}, \citenamefont {Song}, \citenamefont {Troyer},\ and\
  \citenamefont {Soluyanov}}]{WU2017}%
  \BibitemOpen
  \bibfield  {author} {\bibinfo {author} {\bibfnamefont {QuanSheng}\
  \bibnamefont {Wu}}, \bibinfo {author} {\bibfnamefont {ShengNan}\ \bibnamefont
  {Zhang}}, \bibinfo {author} {\bibfnamefont {Hai-Feng}\ \bibnamefont {Song}},
  \bibinfo {author} {\bibfnamefont {Matthias}\ \bibnamefont {Troyer}}, \ and\
  \bibinfo {author} {\bibfnamefont {Alexey~A.}\ \bibnamefont {Soluyanov}},\
  }\bibfield  {title} {\enquote {\bibinfo {title} {Wanniertools : An
  open-source software package for novel topological materials},}\ }\href
  {\doibase https://doi.org/10.1016/j.cpc.2017.09.033} {\bibfield  {journal}
  {\bibinfo  {journal} {Comput. Phys. Commun.}\ }\textbf {\bibinfo {volume}
  {224}},\ \bibinfo {pages} {405 -- 416} (\bibinfo {year} {2018})}\BibitemShut
  {NoStop}%
\bibitem [{\citenamefont {Kresse}\ and\ \citenamefont
  {Furthm\"uller}(1996)}]{PhysRevB.54.11169}%
  \BibitemOpen
  \bibfield  {author} {\bibinfo {author} {\bibfnamefont {G.}~\bibnamefont
  {Kresse}}\ and\ \bibinfo {author} {\bibfnamefont {J.}~\bibnamefont
  {Furthm\"uller}},\ }\bibfield  {title} {\enquote {\bibinfo {title} {Efficient
  iterative schemes for ab initio total-energy calculations using a plane-wave
  basis set},}\ }\href {\doibase 10.1103/PhysRevB.54.11169} {\bibfield
  {journal} {\bibinfo  {journal} {Phys. Rev. B}\ }\textbf {\bibinfo {volume}
  {54}},\ \bibinfo {pages} {11169--11186} (\bibinfo {year} {1996})}\BibitemShut
  {NoStop}%
\bibitem [{Note1()}]{Note1}%
  \BibitemOpen
  \bibinfo {note} {{Both the Quantum Espresso and VASP codes were used for
  calculating structural and electronic properties. Two sets of results were
  found to be in excellent agreement.}}\BibitemShut {Stop}%
\bibitem [{\citenamefont {Cabral}\ \emph {et~al.}()\citenamefont {Cabral},
  \citenamefont {Galbiatti}, \citenamefont {Kwitko-Ribeiro},\ and\
  \citenamefont {Lehmann}}]{HgPt2Se3_structure_1}%
  \BibitemOpen
  \bibfield  {author} {\bibinfo {author} {\bibfnamefont {A.~R.}\ \bibnamefont
  {Cabral}}, \bibinfo {author} {\bibfnamefont {H.~F.}\ \bibnamefont
  {Galbiatti}}, \bibinfo {author} {\bibfnamefont {R.}~\bibnamefont
  {Kwitko-Ribeiro}}, \ and\ \bibinfo {author} {\bibfnamefont {B.}~\bibnamefont
  {Lehmann}},\ }\bibfield  {title} {\enquote {\bibinfo {title} {Platinum
  enrichment at low temperatures and related microstructures, with examples of
  hongshiite (ptcu) and empirical ‘pt2hgse3’ from itabira, minas gerais,
  brazil},}\ }\href {\doibase 10.1111/j.1365-3121.2007.00783.x} {\bibfield
  {journal} {\bibinfo  {journal} {Terra Nova}\ }\textbf {\bibinfo {volume}
  {20}},\ \bibinfo {pages} {32--37}}\BibitemShut {NoStop}%
\bibitem [{\citenamefont {Vymazalová}\ \emph {et~al.}(2012)\citenamefont
  {Vymazalová}, \citenamefont {Laufek}, \citenamefont {Drábek}, \citenamefont
  {Cabral}, \citenamefont {Haloda}, \citenamefont {Sidorinová}, \citenamefont
  {Lehmann}, \citenamefont {Galbiatti},\ and\ \citenamefont
  {Drahokoupil}}]{HgPt2Se3_structure_2}%
  \BibitemOpen
  \bibfield  {author} {\bibinfo {author} {\bibfnamefont {Anna}\ \bibnamefont
  {Vymazalová}}, \bibinfo {author} {\bibfnamefont {František}\ \bibnamefont
  {Laufek}}, \bibinfo {author} {\bibfnamefont {Milan}\ \bibnamefont {Drábek}},
  \bibinfo {author} {\bibfnamefont {Alexandre~Raphael}\ \bibnamefont {Cabral}},
  \bibinfo {author} {\bibfnamefont {Jakub}\ \bibnamefont {Haloda}}, \bibinfo
  {author} {\bibfnamefont {Tamara}\ \bibnamefont {Sidorinová}}, \bibinfo
  {author} {\bibfnamefont {Bernd}\ \bibnamefont {Lehmann}}, \bibinfo {author}
  {\bibfnamefont {Henry~Francisco}\ \bibnamefont {Galbiatti}}, \ and\ \bibinfo
  {author} {\bibfnamefont {Jan}\ \bibnamefont {Drahokoupil}},\ }\bibfield
  {title} {\enquote {\bibinfo {title} {Jacutingaite, pt2hgse3, a new
  platinum-group mineral species from the cauÊ iron-ore deposit, itabira
  district, minas gerais, brazil},}\ }\href {\doibase 10.3749/canmin.50.2.431}
  {\bibfield  {journal} {\bibinfo  {journal} {The Canadian Mineralogist}\
  }\textbf {\bibinfo {volume} {50}},\ \bibinfo {pages} {431} (\bibinfo {year}
  {2012})}\BibitemShut {NoStop}%
\bibitem [{\citenamefont {Slonczewski}\ and\ \citenamefont
  {Weiss}(1958)}]{Graphite_BS}%
  \BibitemOpen
  \bibfield  {author} {\bibinfo {author} {\bibfnamefont {J.~C.}\ \bibnamefont
  {Slonczewski}}\ and\ \bibinfo {author} {\bibfnamefont {P.~R.}\ \bibnamefont
  {Weiss}},\ }\bibfield  {title} {\enquote {\bibinfo {title} {Band structure of
  graphite},}\ }\href {\doibase 10.1103/PhysRev.109.272} {\bibfield  {journal}
  {\bibinfo  {journal} {Phys. Rev.}\ }\textbf {\bibinfo {volume} {109}},\
  \bibinfo {pages} {272--279} (\bibinfo {year} {1958})}\BibitemShut {NoStop}%
\bibitem [{\citenamefont {Lobato}\ and\ \citenamefont
  {Partoens}(2011)}]{AA_graphite}%
  \BibitemOpen
  \bibfield  {author} {\bibinfo {author} {\bibfnamefont {I.}~\bibnamefont
  {Lobato}}\ and\ \bibinfo {author} {\bibfnamefont {B.}~\bibnamefont
  {Partoens}},\ }\bibfield  {title} {\enquote {\bibinfo {title} {Multiple dirac
  particles in aa-stacked graphite and multilayers of graphene},}\ }\href
  {\doibase 10.1103/PhysRevB.83.165429} {\bibfield  {journal} {\bibinfo
  {journal} {Phys. Rev. B}\ }\textbf {\bibinfo {volume} {83}},\ \bibinfo
  {pages} {165429} (\bibinfo {year} {2011})}\BibitemShut {NoStop}%
\bibitem [{\citenamefont {Jin}\ \emph {et~al.}(2019)\citenamefont {Jin},
  \citenamefont {Huang}, \citenamefont {Mei}, \citenamefont {Liu},
  \citenamefont {Lim},\ and\ \citenamefont {Liu}}]{MgB2_nodalline}%
  \BibitemOpen
  \bibfield  {author} {\bibinfo {author} {\bibfnamefont {Kyung-Hwan}\
  \bibnamefont {Jin}}, \bibinfo {author} {\bibfnamefont {Huaqing}\ \bibnamefont
  {Huang}}, \bibinfo {author} {\bibfnamefont {Jia-Wei}\ \bibnamefont {Mei}},
  \bibinfo {author} {\bibfnamefont {Zheng}\ \bibnamefont {Liu}}, \bibinfo
  {author} {\bibfnamefont {Lih-King}\ \bibnamefont {Lim}}, \ and\ \bibinfo
  {author} {\bibfnamefont {Feng}\ \bibnamefont {Liu}},\ }\bibfield  {title}
  {\enquote {\bibinfo {title} {Topological superconducting phase in high-tc
  superconductor mgb2 with dirac-nodal-line fermions},}\ }\href {\doibase
  10.1038/s41524-019-0191-2} {\bibfield  {journal} {\bibinfo  {journal} {Npj
  Comput. Mater.}\ }\textbf {\bibinfo {volume} {5}},\ \bibinfo {pages} {57}
  (\bibinfo {year} {2019})}\BibitemShut {NoStop}%
\bibitem [{\citenamefont {Takane}\ \emph {et~al.}(2018)\citenamefont {Takane},
  \citenamefont {Souma}, \citenamefont {Nakayama}, \citenamefont {Nakamura},
  \citenamefont {Oinuma}, \citenamefont {Hori}, \citenamefont {Horiba},
  \citenamefont {Kumigashira}, \citenamefont {Kimura}, \citenamefont
  {Takahashi},\ and\ \citenamefont {Sato}}]{AlB2_nodalline}%
  \BibitemOpen
  \bibfield  {author} {\bibinfo {author} {\bibfnamefont {Daichi}\ \bibnamefont
  {Takane}}, \bibinfo {author} {\bibfnamefont {Seigo}\ \bibnamefont {Souma}},
  \bibinfo {author} {\bibfnamefont {Kosuke}\ \bibnamefont {Nakayama}}, \bibinfo
  {author} {\bibfnamefont {Takechika}\ \bibnamefont {Nakamura}}, \bibinfo
  {author} {\bibfnamefont {Hikaru}\ \bibnamefont {Oinuma}}, \bibinfo {author}
  {\bibfnamefont {Kentaro}\ \bibnamefont {Hori}}, \bibinfo {author}
  {\bibfnamefont {Kouji}\ \bibnamefont {Horiba}}, \bibinfo {author}
  {\bibfnamefont {Hiroshi}\ \bibnamefont {Kumigashira}}, \bibinfo {author}
  {\bibfnamefont {Noriaki}\ \bibnamefont {Kimura}}, \bibinfo {author}
  {\bibfnamefont {Takashi}\ \bibnamefont {Takahashi}}, \ and\ \bibinfo {author}
  {\bibfnamefont {Takafumi}\ \bibnamefont {Sato}},\ }\bibfield  {title}
  {\enquote {\bibinfo {title} {Observation of a dirac nodal line in
  ${\mathrm{alb}}_{2}$},}\ }\href {\doibase 10.1103/PhysRevB.98.041105}
  {\bibfield  {journal} {\bibinfo  {journal} {Phys. Rev. B}\ }\textbf {\bibinfo
  {volume} {98}},\ \bibinfo {pages} {041105} (\bibinfo {year}
  {2018})}\BibitemShut {NoStop}%
\bibitem [{\citenamefont {Vergniory}\ \emph {et~al.}(2019)\citenamefont
  {Vergniory}, \citenamefont {Elcoro}, \citenamefont {Felser}, \citenamefont
  {Regnault}, \citenamefont {Bernevig},\ and\ \citenamefont
  {Wang}}]{Catalouge_TM}%
  \BibitemOpen
  \bibfield  {author} {\bibinfo {author} {\bibfnamefont {M.~G.}\ \bibnamefont
  {Vergniory}}, \bibinfo {author} {\bibfnamefont {L.}~\bibnamefont {Elcoro}},
  \bibinfo {author} {\bibfnamefont {Claudia}\ \bibnamefont {Felser}}, \bibinfo
  {author} {\bibfnamefont {Nicolas}\ \bibnamefont {Regnault}}, \bibinfo
  {author} {\bibfnamefont {B.~Andrei}\ \bibnamefont {Bernevig}}, \ and\
  \bibinfo {author} {\bibfnamefont {Zhijun}\ \bibnamefont {Wang}},\ }\bibfield
  {title} {\enquote {\bibinfo {title} {A complete catalogue of high-quality
  topological materials},}\ }\href {\doibase 10.1038/s41586-019-0954-4}
  {\bibfield  {journal} {\bibinfo  {journal} {Nature}\ }\textbf {\bibinfo
  {volume} {566}},\ \bibinfo {pages} {480--485} (\bibinfo {year}
  {2019})}\BibitemShut {NoStop}%
\bibitem [{TQC()}]{TQC_website}%
  \BibitemOpen
  \href@noop {} {\enquote {\bibinfo {title}
  {https://www.topologicalquantumchemistry.com},}\ }\bibinfo {howpublished}
  {\url{https://www.topologicalquantumchemistry.com}}\BibitemShut {NoStop}%
\bibitem [{\citenamefont {Tang}\ \emph {et~al.}(2019)\citenamefont {Tang},
  \citenamefont {Po}, \citenamefont {Vishwanath},\ and\ \citenamefont
  {Wan}}]{Viswanath_database}%
  \BibitemOpen
  \bibfield  {author} {\bibinfo {author} {\bibfnamefont {Feng}\ \bibnamefont
  {Tang}}, \bibinfo {author} {\bibfnamefont {Hoi~Chun}\ \bibnamefont {Po}},
  \bibinfo {author} {\bibfnamefont {Ashvin}\ \bibnamefont {Vishwanath}}, \ and\
  \bibinfo {author} {\bibfnamefont {Xiangang}\ \bibnamefont {Wan}},\ }\bibfield
   {title} {\enquote {\bibinfo {title} {Comprehensive search for topological
  materials using symmetry indicators},}\ }\href {\doibase
  10.1038/s41586-019-0937-5} {\bibfield  {journal} {\bibinfo  {journal}
  {Nature}\ }\textbf {\bibinfo {volume} {566}},\ \bibinfo {pages} {486--489}
  (\bibinfo {year} {2019})}\BibitemShut {NoStop}%
\bibitem [{\citenamefont {Teo}\ \emph {et~al.}(2008)\citenamefont {Teo},
  \citenamefont {Fu},\ and\ \citenamefont {Kane}}]{z2_parity}%
  \BibitemOpen
  \bibfield  {author} {\bibinfo {author} {\bibfnamefont {Jeffrey C.~Y.}\
  \bibnamefont {Teo}}, \bibinfo {author} {\bibfnamefont {Liang}\ \bibnamefont
  {Fu}}, \ and\ \bibinfo {author} {\bibfnamefont {C.~L.}\ \bibnamefont
  {Kane}},\ }\bibfield  {title} {\enquote {\bibinfo {title} {Surface states and
  topological invariants in three-dimensional topological insulators:
  Application to ${\text{bi}}_{1\ensuremath{-}x}{\text{sb}}_{x}$},}\ }\href
  {\doibase 10.1103/PhysRevB.78.045426} {\bibfield  {journal} {\bibinfo
  {journal} {Phys. Rev. B}\ }\textbf {\bibinfo {volume} {78}},\ \bibinfo
  {pages} {045426} (\bibinfo {year} {2008})}\BibitemShut {NoStop}%
\bibitem [{Note2()}]{Note2}%
  \BibitemOpen
  \bibinfo {note} {{Note that the topological phase transition to a Dirac
  semimetal state takes place at $\protect \frac {V}{V_0} \sim 0.975$ ($1.031$)
  which corresponds to $2.5\% ~(3.1\%$) decrease (increase) of volume $V_0$.
  This level of strain would be practical to achieve in experiments. We have
  not explored the dynamical stability of the structure under pressure,
  although such a study will be interesting.}}\BibitemShut {Stop}%
\bibitem [{\citenamefont {Politano}\ \emph {et~al.}(2018)\citenamefont
  {Politano}, \citenamefont {Chiarello}, \citenamefont {Ghosh}, \citenamefont
  {Sadhukhan}, \citenamefont {Kuo}, \citenamefont {Lue}, \citenamefont
  {Pellegrini},\ and\ \citenamefont {Agarwal}}]{PtTe2PRL}%
  \BibitemOpen
  \bibfield  {author} {\bibinfo {author} {\bibfnamefont {Antonio}\ \bibnamefont
  {Politano}}, \bibinfo {author} {\bibfnamefont {Gennaro}\ \bibnamefont
  {Chiarello}}, \bibinfo {author} {\bibfnamefont {Barun}\ \bibnamefont
  {Ghosh}}, \bibinfo {author} {\bibfnamefont {Krishanu}\ \bibnamefont
  {Sadhukhan}}, \bibinfo {author} {\bibfnamefont {Chia-Nung}\ \bibnamefont
  {Kuo}}, \bibinfo {author} {\bibfnamefont {Chin~Shan}\ \bibnamefont {Lue}},
  \bibinfo {author} {\bibfnamefont {Vittorio}\ \bibnamefont {Pellegrini}}, \
  and\ \bibinfo {author} {\bibfnamefont {Amit}\ \bibnamefont {Agarwal}},\
  }\bibfield  {title} {\enquote {\bibinfo {title} {3d dirac plasmons in the
  type-ii dirac semimetal ${\mathrm{ptte}}_{2}$},}\ }\href {\doibase
  10.1103/PhysRevLett.121.086804} {\bibfield  {journal} {\bibinfo  {journal}
  {Phys. Rev. Lett.}\ }\textbf {\bibinfo {volume} {121}},\ \bibinfo {pages}
  {086804} (\bibinfo {year} {2018})}\BibitemShut {NoStop}%
\bibitem [{\citenamefont {Facio}\ \emph {et~al.}(2019)\citenamefont {Facio},
  \citenamefont {Das}, \citenamefont {Zhang}, \citenamefont {Koepernik},
  \citenamefont {van~den Brink},\ and\ \citenamefont
  {Fulga}}]{HgPt2Se3_bulk_theory}%
  \BibitemOpen
  \bibfield  {author} {\bibinfo {author} {\bibfnamefont {Jorge~I.}\
  \bibnamefont {Facio}}, \bibinfo {author} {\bibfnamefont {Sanjib~Kumar}\
  \bibnamefont {Das}}, \bibinfo {author} {\bibfnamefont {Yang}\ \bibnamefont
  {Zhang}}, \bibinfo {author} {\bibfnamefont {Klaus}\ \bibnamefont
  {Koepernik}}, \bibinfo {author} {\bibfnamefont {Jeroen}\ \bibnamefont
  {van~den Brink}}, \ and\ \bibinfo {author} {\bibfnamefont {Ion~Cosma}\
  \bibnamefont {Fulga}},\ }\bibfield  {title} {\enquote {\bibinfo {title} {Dual
  topology in jacutingaite ${\mathrm{pt}}_{2}{\mathrm{hgse}}_{3}$},}\ }\href
  {\doibase 10.1103/PhysRevMaterials.3.074202} {\bibfield  {journal} {\bibinfo
  {journal} {Phys. Rev. Materials}\ }\textbf {\bibinfo {volume} {3}},\ \bibinfo
  {pages} {074202} (\bibinfo {year} {2019})}\BibitemShut {NoStop}%
\bibitem [{\citenamefont {Cucchi}\ \emph {et~al.}(2019)\citenamefont {Cucchi},
  \citenamefont {Marrazzo}, \citenamefont {Cappelli}, \citenamefont {Ricco},
  \citenamefont {Bruno}, \citenamefont {Lisi}, \citenamefont {Hoesch},
  \citenamefont {Kim}, \citenamefont {Cacho}, \citenamefont {Besnard},
  \citenamefont {Giannini}, \citenamefont {Marzari}, \citenamefont {Gibertini},
  \citenamefont {Baumberger},\ and\ \citenamefont
  {Tamai}}]{HgPt2Se3_expt_ARPES}%
  \BibitemOpen
  \bibfield  {author} {\bibinfo {author} {\bibfnamefont {I.}~\bibnamefont
  {Cucchi}}, \bibinfo {author} {\bibfnamefont {A.}~\bibnamefont {Marrazzo}},
  \bibinfo {author} {\bibfnamefont {E.}~\bibnamefont {Cappelli}}, \bibinfo
  {author} {\bibfnamefont {S.}~\bibnamefont {Ricco}}, \bibinfo {author}
  {\bibfnamefont {F.~Y.}\ \bibnamefont {Bruno}}, \bibinfo {author}
  {\bibfnamefont {S.}~\bibnamefont {Lisi}}, \bibinfo {author} {\bibfnamefont
  {M.}~\bibnamefont {Hoesch}}, \bibinfo {author} {\bibfnamefont {T.~K.}\
  \bibnamefont {Kim}}, \bibinfo {author} {\bibfnamefont {C.}~\bibnamefont
  {Cacho}}, \bibinfo {author} {\bibfnamefont {C.}~\bibnamefont {Besnard}},
  \bibinfo {author} {\bibfnamefont {E.}~\bibnamefont {Giannini}}, \bibinfo
  {author} {\bibfnamefont {N.}~\bibnamefont {Marzari}}, \bibinfo {author}
  {\bibfnamefont {M.}~\bibnamefont {Gibertini}}, \bibinfo {author}
  {\bibfnamefont {F.}~\bibnamefont {Baumberger}}, \ and\ \bibinfo {author}
  {\bibfnamefont {A.}~\bibnamefont {Tamai}},\ }\bibfield  {title} {\enquote
  {\bibinfo {title} {Bulk and surface electronic structure of the dual-topology
  semimetal pt2hgse3},}\ }\href@noop {} {\ ,\ \bibinfo {eid} {arXiv:1909.05051}
  (\bibinfo {year} {2019})},\ \Eprint {http://arxiv.org/abs/1909.05051}
  {1909.05051} \BibitemShut {NoStop}%
\end{thebibliography}%

\end{document}